# Updated Lagrangian unsaturated periporomechanics for extreme large deformation in unsaturated porous media

Shashank Menon, Xiaoyu Song*

*Department of Civil and Coastal Engineering, University of Florida, Gainesville, FL-32611.*

**Abstract**

Unsaturated periporomechanics is a strong nonlocal poromechanics based on peridynamic state and effective force concept. In the previous periporomechnics the total Lagrangian formulation is adopted for the solid skeleton of porous media. In this article as a new contribution we formulate and implement an updated Lagrangian unsaturated periporomechanics framework for modeling extreme large deformation in unsaturated soils under drained condition, e.g., soil column collapse. In this new framework the so-called bond-associated sub-horizon concept is utilized to enhance the stability and accuracy at extreme large deformation of solid skeleton. The stabilized nonlocal velocity gradient in the deformed configuration is used to update the effective force state from a critical state based visco-plastic model for unsaturated soils. The updated Lagrangian periporomechanics is implemented numerically through an explicit Newmark scheme for high-performance computing. Numerical examples are presented to demonstrate the efficacy of the updated Lagrangian periporomechanics and its robustness in modeling unsaturated soil column collapse under drained condition.

*Keywords:* periporomechanics, unsaturated porous media, peridynamics, extreme large deformation, stabilization, updated Lagrangian

## 1. Introduction

Unsaturated periporomechanics (e.g., [1–4], and a brief introduction in Section 2.1) is a strong nonlocal formulation of classical poromechanics (e.g., [5–7]) based on peridynamic states [8, 9] and effective force state concept (e.g.,[10]). In periporomechnics the solid skeleton of porous media is described by the total Lagrangian framework [1, 11, 12], following the lines in the original peridynamics for solids (e.g., [8, 9, 13, 14]). The total Lagrangian framework for solid mechanics and poromechanics could become unreliable if extreme large deformation occurs (e.g.,[5, 15–17]). For peridynamics for solids, a few studies have been focused on an updated Lagrangian or Eulerian formulation (e.g., [17, 18]). Bergel and Li [17] proposed an updated Lagrangian peridynamic model for solids in which the horizon of a material point is updated in the deformed configuration. The influence function [9] is determined in terms of the relative position of material points in the deformed configuration. The deformation gradient operator and shape tensor are reformulated in the updated Lagrangian framework. Silling et al. [18] presented a Eulerian peridynamic model that define bond forces based on only the current configuration. In [18] the authors demonstrated the thermodynamic consistency of the formulation and the efficacy of the formulation in modeling shock waves and fluid motion. While the Eulerian peridynamic formulation in [18] is robust in the aforementioned applications, it is difficult to simulate history-dependent deformation, e.g., plasticity or viscoplaticity. In this article to realistically simulate the mechanical behavior of unsaturated porous media at extreme large deformation (e.g., unsaturated soil column collapse) we formulate an updated Lagrangian unsaturated periporomechanics framework under drained conditions (i.e., constant negative pore water pressure or matric suction). In this updated Lagrangian periporomechanics framework, it is hypothesized that the horizon of a mixed material point is the same uniform sphere as in the referenced and current/deformed configurations of solid skeleton. In line with this hypothesis, the internal variables of the plastic or viscoplasticity model are stored on Lagrangian material points. In the new framework, the multiphase correspondence constitutive principle in periporomechanics (e.g., [2, 10, 12]) is reformulated in the updated Lagrangian framework, which will be discussed in what follows.

*Corresponding author
*Email address:* xysong@ufl.edu (Xiaoyu Song )

It is known that both the original single-phase (i.e., solids) and multiphase (i.e., porous media) peridynamic correspondence constitutive models exhibit zero-energy deformation mode instability (e.g., [4, 9, 12, 19]). Silling [19] showed that the numerical oscillation in peridynamic correspondence constitutive models for solids is a material instability instead of a pure numerical instability related to the spatial meshless discretization [20]. The reasons of instability and numerical oscillations are (i) the weak dependence of the force state in a bond on its own deformation and (ii) the loss of the non-uniform deformation due to the integration over the whole horizon of a material point. In [12] the authors demonstrated that the multiphase correspondence principle in periporomechanics inherits the zero-energy mode instability in the original peridynamic correspondence principle [9]. We note that numerous techniques have been proposed to circumvent the instability of the original peridynamic correspondence constitutive model for solids under extreme large deformation and/or dynamic loading (e.g., [19, 21–25, 25–28], among others). One technique that could eliminate the instability of peridynamic correspondence material models is the so-called sub-horizon or bond-associated peridynamic correspondence material model [25, 26, 29]. In [29], the authors proposed to decompose the spherically symmetric horizon into discrete sub-horizons that would effectively disturb the radial symmetry of peridynamics. As a material point, each sub-horizon has its own deformation gradient that can remove the smoothing effect of assembling the deformation gradient over the entire horizon. Chen [26] formulated a bond-associated peridynamics for correspondence mate- rial models for solids which is consistent with the sub-horizon concept. Different from the work in [29] the sub-horizons are associated with individual bonds and include only the neighboring material points that are around the bond (i.e., sub-horizon). Gu et al. [25] enhanced the bond- associated sub-horizon peridynamics using the higher-order deformation gradient to improve its accuracy for modeling solids. Note that the sub-horizon or bond-associated peridynamics was proposed for stabilizing peridynamic correspondence constitutive models for single-phase solids. Recently the authors [12] formulated a stabilized multiphase correspondence principle for unsaturated periporomechanics in the total Lagrangian framework.

In this study, as a new contribution, we demonstrate that the updated Lagrangian periporomechanics formulation inherits the stability of the original multiphase correspondence principle in the total Lagrangian formulation of solid skeleton in periporomechanics. Then we implement the bond-associated sub-horizon concept to stabilize the formulated updated Lagrangian periporomechanics. The updated Lagrangian periporomechanics is implemented numerically through an explicit Newmark scheme with Open MPI [30] for high-performance computing. Numerical examples are presented to demonstrate the efficacy of the updated Lagrangian periporomechanics and its robustness in modeling unsaturated soil column collapse under drained conditions. We model unsaturated soil column collapse as an application of the proposed stabilized updated Lagrangian periporomechanics framework in that the soil column collapse is of great relevance in a number of geological and industrial processes such as debris flows [31, 32], landslides [33] and pyroclastic flows [34, 35]. For instance, debris flows during avalanches and mudslides are major geohazards on Earth because debris can travel extensive distances and destroy civil infrastructure such as buildings and roadways. We refer to the literature for numerous studies of soil column collapses under gravity load through physical testing and numerical modeling (e.g., [36–41], among many others).

We note that the periporomechanics model is computationally more demanding than other continuum-based computational methods (e.g., the finite element method) for modeling the mechanics and physics of porous media. We refer to the literature for coupling peridynamics with the finite element method (e.g., [14, 42, 43]) for porous materials. In summary, the new contributions of this article include (i) the formulation of an updated Lagrangian periporomechanics model for extreme large deformation in unsaturated porous media under drained conditions (i.e., uncoupled), (ii) the proof of the lack of stability of the multiphase constitutive correspondence principle in the new framework and the stabilization through the sub-horizon concept, (iii) the numerical implementation for high-performance computing through an explicit Newmark scheme, and (iv) the demonstration of efficacy of the numerical model in modeling unsaturated soil column collapse under drained conditions. For sign convention, the assumption in continuum mechanics is followed, i.e., for the solid skeleton, tensile force/stress is positive and compression is negative, and for fluid pressure compression is positive and tension is negative.

## 2. Updated Lagrangian unsaturated periporomechanics

Periporomechanics has been formulated based on the total Lagrangian approach [10, 44] in which the deformation of solid skeleton is referred to the reference/unreformed configuration of skeleton and the fluid flow is described through the relative Eulerian framework referring to the skeleton in the current configuration. In periporomechanics, it is hypothesized that a porous media body can be conceptualized as a collection of a finite number of mixed material points with two kinds of degree of freedom, i.e., displacement and fluid pressure. A material point $\boldsymbol{X}$ has poromechanical and physical interactions with any material point $\boldsymbol{X}'$ within its neighborhood, $\mathcal{H}$. Here $\mathcal{H}$ is a spherical domain around $\boldsymbol{X}$ with radius $\delta$, i.e., the horizon for the porous medium, in the initial configuration. A stabilized multiphase constitutive correspondence principle (e.g., [10, 12]) has been proposed to incorporate the classical constitutive models for unsaturated soils and physical laws for unsaturated fluid flow in porous media. In this section we reformulate total Lagrangian periporomechanics using the updated Lagrangian framework for extreme large deformation in unsaturated porous media under drained conditions (i.e., constant matric suction).

### 2.1. Updated Lagrangian formulation

For the updated Lagrangian periporomechanics, the equation of motion of a porous body is formulated referring to the current configuration instead of the initial configuration of the same porous body. Figure 1 schematically represent 3 configurations of a porous material body, i.e., the initial/undeformed configuration, current configuration and next configuration following the current configuration. For conciseness of notations, in the current configuration the peridynamic state variable without a prime denotes the variable evaluated at $\boldsymbol{x}$ on the associated bond $\boldsymbol{x}' - \boldsymbol{x}$ and the peridynamic state variable with a prime stands for the variable evaluated at $\boldsymbol{x}'$ on the associated bond $\boldsymbol{x} - \boldsymbol{x}'$, e.g., $\underline{T} = \underline{T}[\boldsymbol{x}]\langle \boldsymbol{x}' - \boldsymbol{x} \rangle$ and $\underline{T}' = \underline{T}[\boldsymbol{x}']\langle \boldsymbol{x} - \boldsymbol{x}' \rangle$. The spatial positions of materials points $\boldsymbol{X}$ and $\boldsymbol{X}'$ in the initial/undeformed configuration are denoted by $\boldsymbol{x}$ and $\boldsymbol{x}'$ in the current/deformed configuration, respectively. Let $\boldsymbol{u}$ and $\boldsymbol{u}'$ be the displacement vectors of $\boldsymbol{x}$ and $\boldsymbol{x}'$ referring to the current configuration. The bond between $\boldsymbol{x}$ and $\boldsymbol{x}'$ is denoted by $\boldsymbol{\zeta}$, i.e., $\boldsymbol{\zeta} = \boldsymbol{x}' - \boldsymbol{x}$ . For the updated Lagrangian periporomechanics formulation, it is hypothesized that the horizon variable $\delta$ is constant. In this sense, the family $\mathscr{H}$ (i.e., horizon) of a material point $\boldsymbol{x}$ in the current configuration is defined as

$$\mathscr{H} := \{\boldsymbol{x}' | \boldsymbol{x}' \in \mathscr{B}, 0 \leq |\boldsymbol{\zeta}| \leq \delta\}, \tag{1}$$

where $\mathscr{B}$ denotes a porous media body. Note that this hypothesis is consistent with the Eulerian formulation of peridynamics for solids in [18]. However, the material points of solid skeleton is described by their motions. In this sense, with deformation, the set of material points in the horizon of material point $\boldsymbol{x}$ can be different from time to time in extreme large deformation regime. Therefore, with this hypothesis the extreme distortion of the horizon for extreme large deformation of the solid skeleton in the total Lagrangian formulation can be avoided. As in the total Lagrangian formulation, in the updated Lagrangian formulation the total force vector state at material point $\boldsymbol{x}$ along bond $\boldsymbol{\zeta}$ under the assumption of passive pore air pressure can be decomposed into

$$\underline{\mathscr{T}} = \overline{\underline{\mathscr{T}}} - S_r \underline{\mathscr{T}}_w, \tag{2}$$

where $\overline{\mathscr{T}}$ and $\mathscr{T}_w$ are the effective force state of the porous media and the water force states, respectively, and $S_r$ is the degree of saturation in the current configuration. Assuming weightless pore air the density of the mixture $\rho$ can be written as

$$\rho = \rho_s(1-\phi) + S_r \rho_w \phi, \tag{3}$$

where $\rho_s$ is the intrinsic density of the solid and $\rho_w$ is the intrinsic density of water, and $\varphi$ is the porosity in the current configuration.

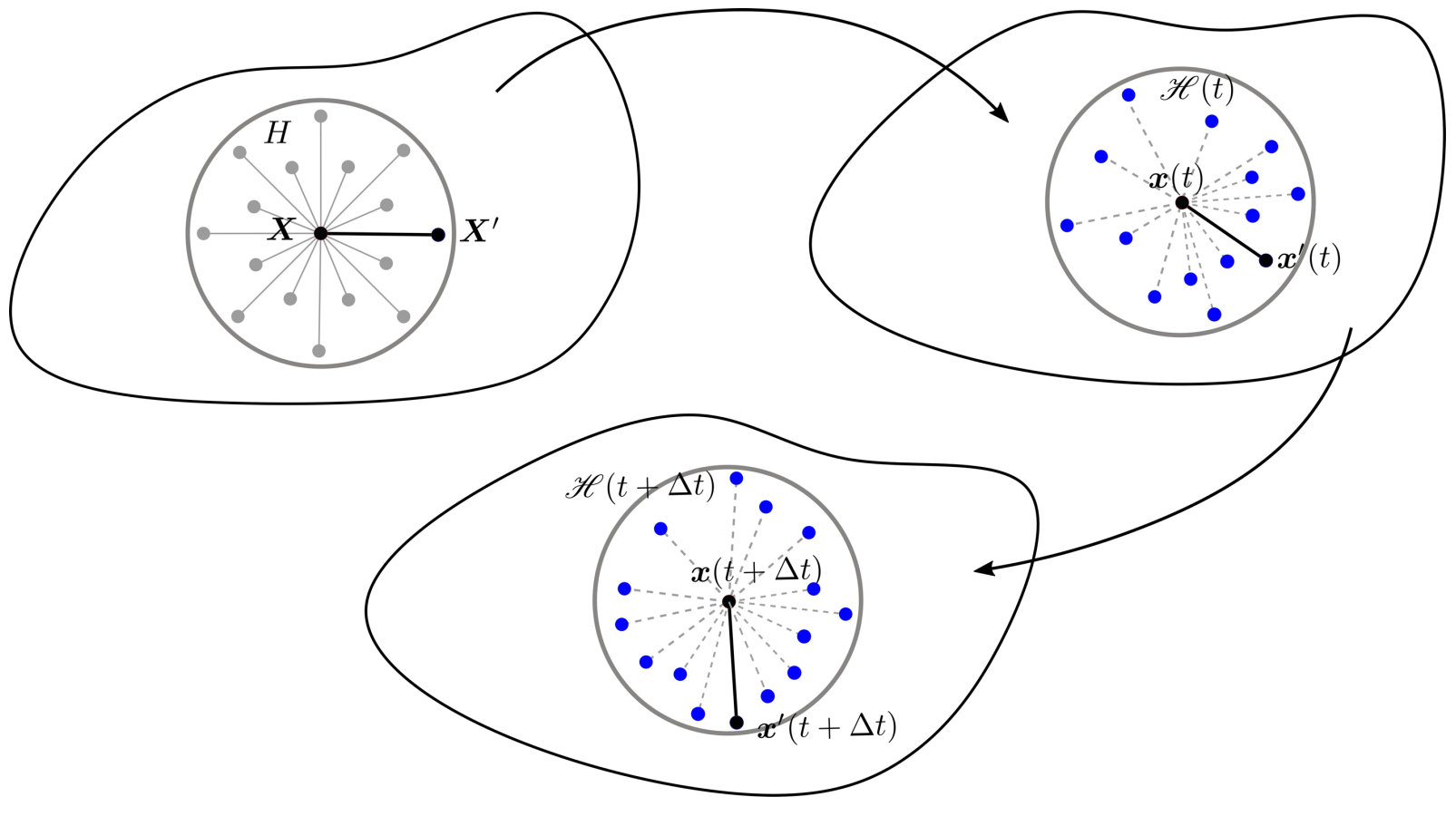

Figure 1: Schematic depiction of the kinematics of the solid skeleton in the updated Lagrangian formulation: initial configuration, current configuration (reference), and next configuration.

Let $\boldsymbol{u}$ be the displacement vector of material point $\boldsymbol{x}$ in the current configuration. Following the lines in the total Lagrangian periporomechanics [3], the equation of motion of the updated Lagrangian periporomechanics can be written as

$$\rho \ddot{\boldsymbol{u}} = \int_{\mathscr{H}} \left( \underline{\mathscr{T}} [\boldsymbol{x}] \langle \underline{\boldsymbol{\zeta}} \rangle - \underline{\mathscr{T}}' [\boldsymbol{x}'] \langle \underline{\boldsymbol{\zeta}'} \rangle \right) \mathrm{d}\mathscr{V}' + \rho \boldsymbol{g}, \tag{4}$$

where $\underline{\mathscr{T}} [\boldsymbol{x}] \langle \underline{\boldsymbol{\zeta}} \rangle$ and $\underline{\mathscr{T}}' [\boldsymbol{x}'] \langle \underline{\boldsymbol{\zeta}'} \rangle$ are the total force vector state at material points $\boldsymbol{x}$ and $\boldsymbol{x}'$ in the current configuration, respectively, $\ddot{\boldsymbol{u}}$ is the acceleration, and $\boldsymbol{g}$ is gravity acceleration vector.

Following [9], the spatial shape tensor $\mathscr{K}$ in the current configuration can be defined as

$$\mathscr{K} = \int_{\mathscr{H}} \underline{\omega} \langle \underline{\boldsymbol{\zeta}} \rangle \, \underline{\boldsymbol{\zeta}} \otimes \underline{\boldsymbol{\zeta}} \, \mathrm{d}\mathscr{V}', \tag{5}$$

where $\underline{\omega} \langle \underline{\boldsymbol{\zeta}} \rangle$ is an influence function. Then it follows from the notion of original reduction operator [9, 14] (i.e., referring to the initial configuration), we can define the spatial gradient operator [17] in the current configuration as

$$\mathscr{G}(\boldsymbol{z}) = \left[ \int_{\mathscr{H}} \underline{\omega} \langle \underline{\boldsymbol{\zeta}} \rangle (\boldsymbol{z}' - \boldsymbol{z}) \otimes \underline{\boldsymbol{\zeta}} \, \mathrm{d}\mathscr{V}' \right] \mathscr{K}^{-1} \tag{6}$$

where $\boldsymbol{z}'$ and $\boldsymbol{z}$ are vector variables at material points $\boldsymbol{x}'$ and $\boldsymbol{x}$, respectively. Using (6) we can define the velocity gradient $\mathscr{L}$. Recall from nonlinear continuum mechanics that $\mathscr{L}$ as the spatial gradient of the velocity vector $\dot{\boldsymbol{u}}$ reads

$$\mathscr{L} = \frac{\partial \dot{\boldsymbol{u}}}{\partial \boldsymbol{x}}. \tag{7}$$

Next, it follows from (6) and (7) we obtain the nonlocal velocity gradient as

$$\mathscr{L} = \mathscr{G}(\dot{\boldsymbol{u}}) = \left( \int_{\mathscr{H}} \underline{\omega} \, \underline{\dot{\mathscr{Y}}} \otimes \underline{\boldsymbol{\zeta}} \, \mathrm{d}\mathscr{V}' \right) \mathscr{K}^{-1}, \tag{8}$$

Given (8), the rate of nonlocal deformation can be readily obtained as

$$\mathscr{D} = \frac{1}{2}[\mathscr{L} + \mathscr{L}^T]. \tag{9}$$

The rate of deformation tensor can be used to determine the effective Cauchy stress tensor $\overline{\boldsymbol{\sigma}}$ through a classical constitutive model for unsaturated soils (e.g., [1, 45]). Then, the rate form of the strain energy of the solid skeleton of an unsaturated porous material body $B$ under pure elastic deformation reads [9, 10],

$$\begin{aligned}
\dot{\mathscr{W}} &= \int_{\mathscr{B}} \overline{\sigma}_{ij} \mathscr{D}_{ij} \, \mathrm{d}\mathscr{V} \\
&= \int_{\mathscr{B}} \overline{\sigma}_{ij} \mathscr{L}_{ij} \, \mathrm{d}\mathscr{V} \\
&= \int_{\mathscr{B}} \overline{\sigma}_{ij} \left( \int_{\mathscr{H}} \underline{\omega} \dot{\underline{\mathscr{Y}}}_i \underline{\zeta}_p \, \mathrm{d}\mathscr{V}' \right) \mathscr{K}_{pj}^{-1} \, \mathrm{d}\mathscr{V} \\
&= \int_{\mathscr{B}} \left( \int_{\mathscr{H}} \underline{\omega} \dot{\underline{\mathscr{Y}}}_i \underline{\zeta}_p \, \mathrm{d}\mathscr{V}' \right) \mathscr{K}_{pj}^{-1} \overline{\sigma}_{ji} \, \mathrm{d}\mathscr{V} \\
&= \int_{\mathscr{B}} \int_{\mathscr{B}} \underline{\omega} \underline{\zeta}_p \mathscr{K}_{pj}^{-1} \overline{\sigma}_{ji} \dot{\underline{\mathscr{Y}}}_i \, \mathrm{d}\mathscr{V}' \, \mathrm{d}\mathscr{V},
\end{aligned} \tag{10}$$

where $i, j, p = 1, 2, 3$. The rate of strain energy of the solid skeleton of porous media underpure elastic deformation referring to the current configuration reads [10],

$$\dot{\mathscr{W}} = \int_{\mathscr{B}} \int_{\mathscr{B}} \overline{\underline{\mathscr{T}}}_i \, \dot{\underline{\mathscr{Y}}}_i \, \mathrm{d}\mathscr{V}' \mathrm{d}\mathscr{V}. \tag{11}$$

It follows from (10) and (11) that the effective force state can be expressed as

$$\overline{\underline{\mathscr{T}}} = \underline{\omega} \underline{\boldsymbol{\zeta}} \mathscr{K}^{-1} \overline{\boldsymbol{\sigma}}. \tag{12}$$

Through the effective force state concept (see equation 2) the fluid force state can be expressed as

$$\underline{\mathscr{T}}_w = \underline{\omega} \underline{\boldsymbol{\zeta}} \mathscr{K}^{-1} (p_w \mathbf{1}). \tag{13}$$

Finally, substituting (12), (13) and (2) into (4) the motion of equation can be written as

$$\rho \ddot{\mathbf{u}} = \int_{\mathscr{H}} \left[ \underline{\omega} \underline{\boldsymbol{\zeta}} \mathscr{K}^{-1} (\overline{\boldsymbol{\sigma}} - S_r p_w \mathbf{1}) - \underline{\omega}' \, \underline{\boldsymbol{\zeta}}' \mathscr{K}'^{-1} (\overline{\boldsymbol{\sigma}}' - S_r' p_w' \mathbf{1}) \right] \, \mathrm{d}\mathscr{V}' + \rho \boldsymbol{g}. \tag{14}$$

The degree of saturation $S_r$ can be determined from the soil-water retention curve (e.g., [46, 47]) that depends on the volume strain of the solid skeleton (e.g., porosity). In this study, we adopt the one in [15, 16, 48] which reads

$$S_r(\mathscr{J}, \phi, p_w) = \left\{ 1 + \left[ -a_1 \left( \frac{\mathscr{J}}{1 - \phi} - 1 \right)^{a_2} \mathscr{J} p_w \right]^n \right\}^{(n-1)/n}, \tag{15}$$

where $a_1$, $a_2$, and $n$ are all material parameters. The evolution of porosity can be written as

$$\phi(\mathscr{J}) = 1 - \frac{(1 - \phi)}{\mathscr{J}}, \tag{16}$$

where $\mathscr{J}$ is the determinant of the spatial deformation gradient $\mathscr{F}$. From (6), the spatial deformation gradient can be written as

$$\mathscr{F} = \mathscr{G} = \left[ \int_{\mathscr{H}} \underline{\omega} \, \underline{\mathscr{Y}} \otimes \underline{\boldsymbol{\zeta}} \, \mathrm{d}\mathscr{V}' \right] \mathscr{K}^{-1}. \tag{17}$$

It can be proved as follows. Given that the spatial deformation gradient $\mathscr{F}$ maps $\boldsymbol{\zeta}$ onto $\boldsymbol{\mathscr{Y}}$,

$$\underline{\mathscr{Y}} = \mathscr{F}\underline{\boldsymbol{\zeta}}. \tag{18}$$

Substituting (18) into (17)

$$\begin{aligned} \mathscr{G}(\boldsymbol{x}) &= \int_{\mathscr{H}} \underline{\omega \mathscr{Y}} \otimes \underline{\boldsymbol{\zeta}} \, \mathrm{d}\mathscr{V}' \left[ \int_{\mathscr{H}} \underline{\omega} \, \underline{\boldsymbol{\zeta}} \otimes \underline{\boldsymbol{\zeta}} \, \mathrm{d}\mathscr{V}' \right]^{-1}, \\ &= \int_{\mathscr{H}} \underline{\omega} (\mathscr{F} \underline{\boldsymbol{\zeta}}) \otimes \underline{\boldsymbol{\zeta}} \, \mathrm{d}\mathscr{V}' \left[ \int_{\mathscr{H}} \underline{\omega} \, \underline{\boldsymbol{\zeta}} \otimes \underline{\boldsymbol{\zeta}} \, \mathrm{d}\mathscr{V}' \right]^{-1}, \\ &= \mathscr{F} \int_{\mathscr{H}} \underline{\omega \boldsymbol{\zeta}} \otimes \underline{\boldsymbol{\zeta}} \, \mathrm{d}\mathscr{V}' \left[ \int_{\mathscr{H}} \underline{\omega \boldsymbol{\zeta}} \otimes \underline{\boldsymbol{\zeta}} \, \mathrm{d}\mathscr{V}' \right]^{-1} \\ &= \mathscr{F}. \end{aligned} \tag{19}$$

In what follows, we demonstrate that the nonlocal spatial gradient defined in (17) only represent the uniform deformation state. Thus, the spatial correspondence principle inherits the zero-energy mode instability issue. The nonuniform part of the solid spatial deformation state reads

$$\underline{\mathscr{R}} = \underline{\mathscr{Y}} - \mathscr{F}\underline{\boldsymbol{\zeta}}. \tag{20}$$

Substituting (20) into the nonlocal spatial deformation gradient in (17),

$$\begin{aligned} \boldsymbol{\mathscr{F}}\left(\underline{\mathscr{R}}\right) &= \left( \int_{\mathscr{H}} \underline{\omega \boldsymbol{\mathscr{Y}}} \otimes \underline{\boldsymbol{\zeta}} \, \mathrm{d}\mathscr{V}' \right) \boldsymbol{\mathscr{K}}^{-1} \\ &= \int_{\mathscr{H}} \underline{\omega} \, \underline{\boldsymbol{\mathcal{R}}} \otimes \underline{\boldsymbol{\zeta}} \, \mathrm{d}\mathscr{V}' \left( \int_{\mathscr{H}} \underline{\omega} \, \underline{\boldsymbol{\zeta}} \otimes \underline{\boldsymbol{\zeta}} \, \mathrm{d}\mathscr{V}' \right)^{-1} \\ &= \int_{\mathscr{H}} \underline{\omega} \left( \underline{\boldsymbol{\mathscr{Y}}} - \boldsymbol{\mathscr{F}} \underline{\boldsymbol{\zeta}} \right) \otimes \underline{\boldsymbol{\zeta}} \, \mathrm{d}\mathscr{V}' \left( \int_{\mathscr{H}} \underline{\omega} \, \underline{\boldsymbol{\zeta}} \otimes \underline{\boldsymbol{\zeta}} \, \mathrm{d}\mathscr{V}' \right)^{-1} \\ &= \left( \int_{\mathscr{H}} \underline{\omega} \left( \underline{\boldsymbol{\mathscr{Y}}} \otimes \underline{\boldsymbol{\zeta}} \right) \mathrm{d}\mathscr{V}' - \boldsymbol{\mathscr{F}} \int_{\mathscr{H}} \underline{\omega} \left( \underline{\boldsymbol{\zeta}} \otimes \underline{\boldsymbol{\zeta}} \right) \mathrm{d}\mathscr{V}' \right) \left( \int_{\mathscr{H}} \underline{\omega} \, \underline{\boldsymbol{\zeta}} \otimes \underline{\boldsymbol{\zeta}} \, \mathrm{d}\mathscr{V}' \right)^{-1} \\ &= \boldsymbol{\mathscr{F}} - \boldsymbol{\mathscr{F}} \boldsymbol{\mathscr{K}} \boldsymbol{\mathscr{K}}^{-1} = \mathbf{0}. \end{aligned} \tag{21}$$

It follows from (21) that the nonuniform part of the deformation state is missing in the spatial deformation gradient. Next, we present a stabilization scheme for the spatial corresponding model for the solid skeleton based on the bond-associated sub-horizon concept.

## 2.2. *Stabilization through the sub-horizon based method*

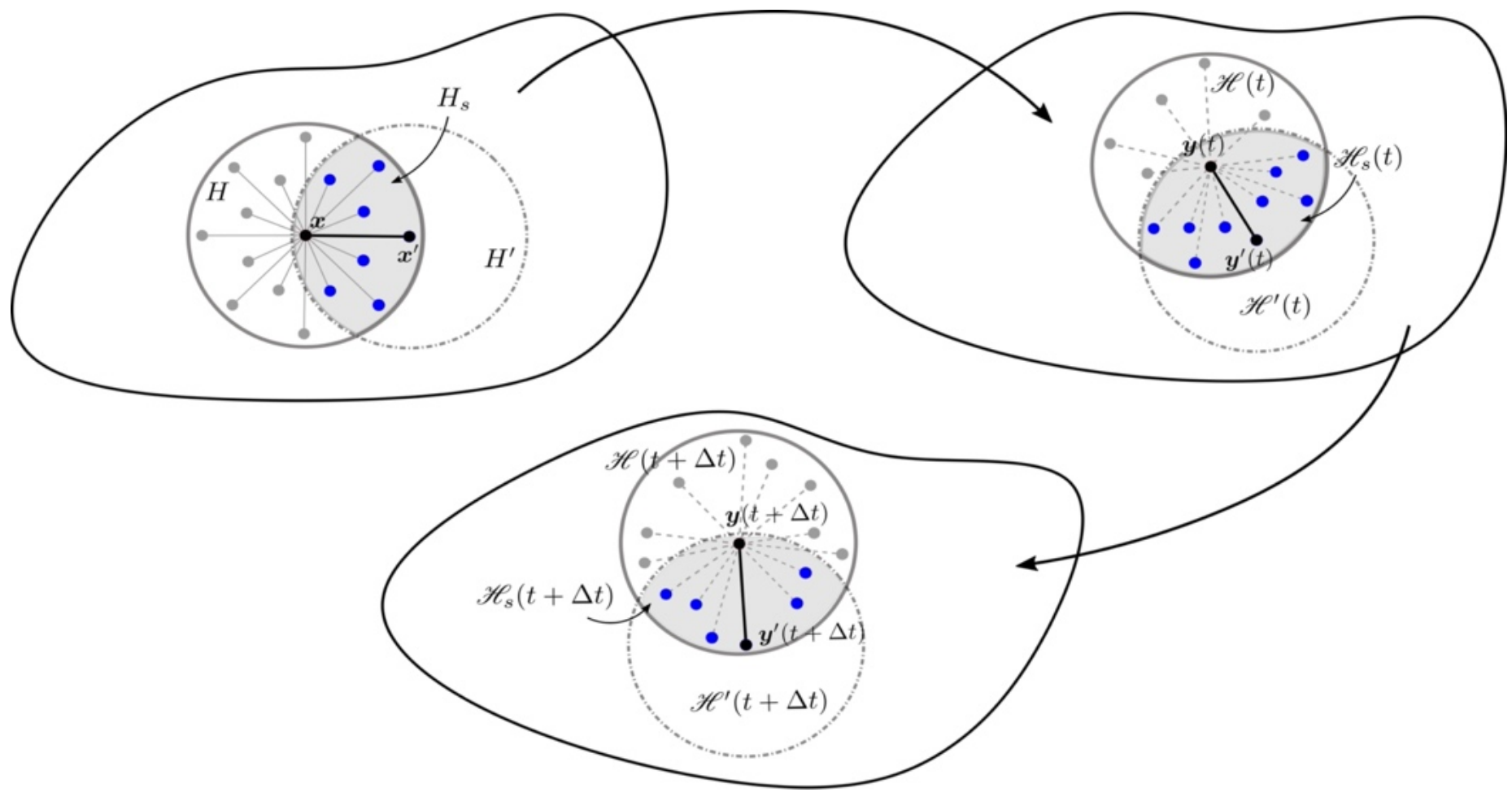

Figure 2: Schematic of the sub-horizon concept for the updated Lagrangian framework.

In this study, we apply the sub-horizon concept (e.g., [25, 26, 29]) to circumvent the zero-energy mode instability of correspondence material models for solid skeleton in the updatedLagrangian periporomechanics. We refer to the literature for other stabilization techniques (e.g., [12, 19, 25]). Figure 2 plots the bond-associated sub-horizon concept both in the ini- tial/reference and current/deformed configurations of a porous material body. As shown in Figure 2, in the horizon of material point $\boldsymbol{x}$ each bond is endorsed with a sub-horizon composed of material points around the bond that is used to determine the nonlocal deformation gradient, the effective stress and the effective force state on that bond. Therefore the sub-horizon stabi- lized scheme can guarantee a non-unique mapping between between the deformation gradient and the force state on individual bond. We note that the original bond-associated sub-horizon formulation was based on the total Lagrangian approach (i.e., the sub-horizon refers to the initial/undeformed configuration). In the updated Lagrangian periporomechanics formulation, this bond-associated sub-horizon concept is reformulated in the current/deformation configu- ration with updated material points in the horizon of a material point and the sub-horizon forindividual bond.
Referring to Figure 2 the sub-horizon $H_s$ for the bond $\boldsymbol{\zeta}$ at material point $\boldsymbol{x}$ in the current configuration of the solid skeleton reads

$$\mathcal{H}_s = \mathcal{H} \cap \mathcal{H}'. \tag{22}$$

It follows from (17), (5) and (22) that the sub-horizon based spatial shape tensor and deformation gradient can be written as

$$\boldsymbol{\mathcal{K}}_s = \int_{\mathcal{H}_s} \underline{\omega}_{\mathcal{Y}} \, \underline{\zeta} \otimes \underline{\zeta} \, \mathrm{d}\mathcal{V}', \tag{23}$$

$$\boldsymbol{\mathcal{F}}_s = \left( \int_{\mathcal{H}_s} \underline{\omega}_{\mathcal{Y}} \underline{\boldsymbol{\mathcal{Y}}} \otimes \underline{\zeta} \, \mathrm{d}\mathcal{V}' \right) \boldsymbol{\mathcal{K}}_s^{-1}. \tag{24}$$

From (8), the sub-horizon based spatial velocity gradient reads,

$$\boldsymbol{\mathcal{L}}_s = \left( \int_{\mathcal{H}_s} \underline{\omega} \, \dot{\underline{\boldsymbol{\mathcal{Y}}}} \otimes \underline{\zeta} \, \mathrm{d}\mathcal{V}' \right) (\boldsymbol{\mathcal{K}}_s)^{-1}, \tag{25}$$

With (25) the sub-horizon based rate of deformation reads

$$\boldsymbol{\mathcal{D}}_s = \frac{1}{2} [\boldsymbol{\mathcal{L}}_s + \boldsymbol{\mathcal{L}}_s^T]. \tag{26}$$

Given (26) the effective Cauchy stress tensor $\bar{\boldsymbol{\sigma}}_s$ can be determined in the current configuration through the classical constitutive model.

Next, we derive the effective force state on a bond $\boldsymbol{\zeta}$ at material point $\boldsymbol{x}$ following the linesin the updated Lagrangian formulation in Section 2.1. The rate of strain energy density at $\boldsymbol{x}$ on the bond $\boldsymbol{\zeta}$ that is determined by the bond-associated sub-horizon method can be written as

$$\dot{\overline{\mathcal{W}}}_s \langle \zeta \rangle = \int_{\mathcal{H}_s} \overline{\underline{\mathcal{T}}}_s \cdot \dot{\underline{\boldsymbol{\mathcal{Y}}}} \, \mathrm{d}\mathcal{V}', \tag{27}$$

where the sub-horizon effective force state on bond $\boldsymbol{\zeta}$ in the current configuration. $\dot{W}_s(\boldsymbol{\zeta})$ can be related to the rate total strain energy density $\dot{\overline{W}}$ at $\boldsymbol{x}$ through a volume fraction factor $\varphi_s$ as

$$\dot{\overline{\mathcal{W}}}_s \langle \zeta \rangle = \varphi_s \dot{\overline{\mathcal{W}}}, \tag{28}$$

with

$$\varphi_s = \frac{\int_{\mathcal{H}_s} 1 \, \mathrm{d}\mathcal{V}'}{\int_{\mathcal{H}} 1 \, \mathrm{d}\mathcal{V}'}. \tag{29}$$

From the local theory the rate of strain energy density at material point $\boldsymbol{x}$ reads

$$
\begin{aligned}
\dot{\overline{\mathscr{W}}} &= \overline{\boldsymbol{\sigma}}_s : \boldsymbol{\mathscr{D}}_s \\
&= \overline{\boldsymbol{\sigma}}_s : \left( \int_{\mathcal{H}_s} \underline{\omega}\ \dot{\underline{\boldsymbol{\mathscr{Y}}}} \otimes \underline{\boldsymbol{\zeta}}\ \mathrm{d}\mathscr{V}' \right) (\boldsymbol{\mathscr{K}}_s)^{-1} \\
&= \int_{\mathcal{H}_s} \underline{\omega}\ \underline{\boldsymbol{\zeta}} (\boldsymbol{\mathscr{K}}_s)^{-1} \overline{\boldsymbol{\sigma}}_s \dot{\underline{\boldsymbol{\mathscr{Y}}}}\ \mathrm{d}\mathscr{V}'.
\end{aligned}
\tag{30}
$$

It follows from (27), (28) and (30) the sub-horizon based effective force state can be expressed as

$$
\overline{\underline{\mathscr{T}}}_s = \varphi_s \underline{\omega}\ \underline{\boldsymbol{\zeta}} (\boldsymbol{\mathscr{K}}_s)^{-1} \overline{\boldsymbol{\sigma}}_s. \tag{31}
$$

From the effective force state concept and (31), the sub-horizon based total force state reads

$$
\underline{\mathscr{T}}_s = \varphi_s \underline{\omega}\ \underline{\boldsymbol{\zeta}} (\boldsymbol{\mathscr{K}}_s)^{-1} (\overline{\boldsymbol{\sigma}}_s - S_r p_w \mathbf{1}). \tag{32}
$$

Then, the equation of motion (14) can be rewritten as

$$
\rho \ddot{\mathbf{u}} = \int_{\mathcal{H}} \left[ \varphi_s \underline{\omega}\ \underline{\boldsymbol{\zeta}} (\boldsymbol{\mathscr{K}}_s)^{-1} (\overline{\boldsymbol{\sigma}}_s - S_r p_w \mathbf{1}) - \varphi'_s \underline{\omega}'\ \underline{\boldsymbol{\zeta}}' (\boldsymbol{\mathscr{K}}'_s)^{-1} (\overline{\boldsymbol{\sigma}}'_s - S'_r p'_w \mathbf{1}) \right] \mathrm{d}\mathscr{V}' + \rho \boldsymbol{g}. \tag{33}
$$

In what follows, we demonstrate that the updated Lagrangian sub-horizon correspondence constitutive model satisfies the sufficient condition (e.g.,[12, 19]) that mitigates the zero-energy mode instability. This criterion in terms of the rate form of the effective force state and deformation state can be written as

$$
\int_{\mathcal{H}} \dot{\overline{\underline{\mathscr{T}}}} \cdot \dot{\underline{\boldsymbol{\mathscr{Y}}}} \mathrm{d}\mathscr{V}' > 0. \tag{34}
$$

It follows from (31) that this condition can be expressed using the sub-horizon based effective force state as

$$
\int_{\mathcal{H}} \left( \int_{\mathcal{H}_s} \dot{\overline{\underline{\mathscr{T}}}} \cdot \dot{\underline{\boldsymbol{\mathscr{Y}}}} \mathrm{d}\mathscr{V}' \right) \mathrm{d}\mathscr{V}' > 0, \tag{35}
$$

where

$$
\begin{aligned}
\dot{\overline{\underline{\mathscr{T}}}}_i &= \frac{1}{\varphi_s} \dot{\overline{\underline{\mathscr{T}}}}_{s,i} \\
&= \underline{\omega}\ \underline{\zeta}_k \mathscr{K}^{-1}_{s,kj} \dot{\overline{\sigma}}_{s,ji}.
\end{aligned}
\tag{36}
$$

Assuming a local elastic material model, we have

$$
\begin{aligned}
\dot{\overline{\sigma}}_{s,ij} &= \mathscr{C}_{ijmn} \mathscr{D}_{s,mn} \\
&= \mathscr{C}_{ijmn} \mathscr{L}_{s,mn},
\end{aligned}
\tag{37}
$$

where $C_{ijmn}$ is the fourth-order elastic stiffness tensor, and $m, n = 1, 2, 3$. Substituting (37) into (35) yields

$$
\begin{aligned}
\int_{\mathcal{H}} \left( \int_{\mathcal{H}_s} \dot{\overline{\underline{\mathscr{T}}}} \cdot \dot{\underline{\boldsymbol{\mathscr{Y}}}} \mathrm{d}\mathscr{V}' \right) \mathrm{d}\mathscr{V}' &= \int_{\mathcal{H}} \left( \int_{\mathcal{H}_s} \left( \underline{\omega}\ \underline{\zeta}_k \mathscr{K}^{-1}_{s,kj} \mathrm{d}\dot{\overline{\sigma}}_{s,ji} \right) \dot{\mathscr{Y}}_i\ \mathrm{d}\mathscr{V}' \right) \mathrm{d}\mathscr{V}' \\
&= \int_{\mathcal{H}} \left( \int_{\mathcal{H}_s} \underline{\omega}\ \underline{\zeta}_k \mathscr{K}^{-1}_{s,kj} \dot{\overline{\sigma}}_{s,ji} \dot{\mathscr{Y}}_i\ \mathrm{d}\mathscr{V}' \right) \mathrm{d}\mathscr{V}' \\
&= \int_{\mathcal{H}} \left( \dot{\overline{\sigma}}_{s,ij} \left( \int_{\mathcal{H}_s} \underline{\omega}\ \dot{\mathscr{Y}}_i\ \underline{\zeta}_k\ \mathrm{d}\mathscr{V}' \right) \mathscr{K}^{-1}_{s,kj} \right) \mathrm{d}\mathscr{V}' \\
&= \int_{\mathcal{H}} \left( \dot{\overline{\sigma}}_{s,ij} \mathscr{L}_{s,ij} \right) \mathrm{d}\mathscr{V}' \\
&= \int_{\mathcal{H}} \left( \mathscr{L}_{s,ij} \mathscr{C}_{ijmn} \mathscr{L}_{s,mn} \right) \mathrm{d}\mathscr{V}' > 0.
\end{aligned}
\tag{38}
$$

It follows from (25) and (38) that (35) holds given $\mathscr{Y} > 0$ (i.e., nonzero increment of deformation state). Therefore, the sub-horizon based correspondence constitutive model is stable in the updated Lagrangian formulation. In this study we assume $\delta = \delta'$ that implies $\mathscr{L}_s = \mathscr{L}'_s$. By this assumption, it will guarantee

$$\underline{\mathscr{T}}_s = -\underline{\mathscr{T}}'_s. \tag{39}$$

Next, we introduce a classical viscoplastic model for unsaturated porous media that will be implemented in the updated Lagrangian periporomechanics framework.

### 2.3. *Constitutive models for unsaturated soil*

In this section, we briefly introduce the key element of a critical state viscoplasticity modelfor unsaturated soils using the Perzyna model of viscoplasticity [49]. For the local viscoplatic constitutive model, the yield function reads

$$f(\bar{p}, q, \bar{p}_c) = \frac{q^2}{M^2} + \bar{p}(\bar{p} - \bar{p}_c), \tag{40}$$

where $\bar{p} = \mathrm{tr}(\overline{\boldsymbol{\sigma}})/3$ is the mean effective stress, $q = \sqrt{3/2}|\overline{\boldsymbol{\sigma}} - \bar{p}\mathbf{1}|$ is the equivalent shear stress, $\overline{\boldsymbol{\sigma}}$ is the effective stress tensor, $M$ is the slope of the critical state line, and $\bar{p}_c$ is the effective pre-consolidation pressure that evolves with the viscoplastic volumetric strain $\varepsilon_v^{vp}$ and matric suction (i.e., $-p_w$ assuming passive pore air pressure. The effective mean stress $\bar{p}$ and the shear stress $q$ can be written as

$$\bar{p} = K\varepsilon_v^e, \quad q = 3\mu_s\varepsilon_s^e, \tag{41}$$

where $K$ and $\mu_s$ are the elastic bulk and shear moduli, respectively, and $\varepsilon_v^e = \mathrm{tr}(\boldsymbol{\varepsilon}^e)$ and $\varepsilon_s^e = \sqrt{\frac{2}{3}}|\boldsymbol{\epsilon}^e - \frac{1}{3}\varepsilon_v^e\mathbf{1}|$ are elastic volumetric strain and shear strain.

The total strain rate is decomposed into

$$\dot{\boldsymbol{\varepsilon}} = \dot{\boldsymbol{\varepsilon}}^e + \dot{\boldsymbol{\varepsilon}}^{vp}. \tag{42}$$

Assuming the Perzyna type of viscoplasticity, the rate of viscoplastic strain tensor $\dot{\boldsymbol{\varepsilon}}^{vp}$ can be expressed as

$$\boldsymbol{\varepsilon}^{\dot{v}p} = \frac{\langle f \rangle}{\eta}\frac{\partial f}{\partial \overline{\boldsymbol{\sigma}}}, \tag{43}$$

where  is the viscosity coefficient and $\langle\rangle$ here are the Macaulay brackets operator

$$\langle f \rangle = \begin{cases} 0 & f \leq 0 \\ f & f > 0. \end{cases} \tag{44}$$

The apparent pre-consolidation pressure $\bar{p}_c$ [1, 50] can be written as

$$\bar{p}_c = -\exp(b_1)(-p_c)^{b_2} \tag{45}$$

where $b_1$ and $b_2$ are variables depending on the degree of saturation and matric suction (see [1, 50] for more details) and $p_c$ reads

$$\dot{p}_c = \frac{-p_c}{\tilde{\lambda} - \tilde{\kappa}}\mathrm{tr}(\dot{\boldsymbol{\epsilon}}^{vp}), \tag{46}$$

where $\tilde{\lambda}$ and $\tilde{\kappa}$ are compression index and swell index, respectively.

## 3. Numerical implementation

### 3.1. *Discretization in space*

The equation of motion (33) is discretized in space by an updated Lagrangian meshfree scheme. In this method, a porous continuum material is discretized into a finite number of mixed material points (i.e., solid skeleton and pore water). Under the assumption of drained condition, each material point has one kind of degree of freedom (i.e., displacement) because the negative pore water pressure (matric suction) is constant at each individual material point under drained condition (i.e., one-way coupling). The uniform grid is used to spatially discretize the problem domain in which all material points have an identical size. Figure 3 provides a flowchart of the global solution procedure and Algorithm 1 summarize the detailed steps for updating the total force states (i.e., effective force and fluid force states) at load step $n+1$ from time step $n$. At time step $n$ the material points in the family of a material point are updated through a search algorithm and each individual neighboring material points are described by their material coordinates (i.e., Lagrangian). For time step $n+1$, calculations of all variables (e.g., sub-horizon, deformation gradient, velocity gradient) are referred to the configuration of solid skeleton at time step $n$ (i.e., updated Lagrangian formulation) instead of the initial/undeformed configuration of solid skeleton (i.e., total Lagrangian formulation).

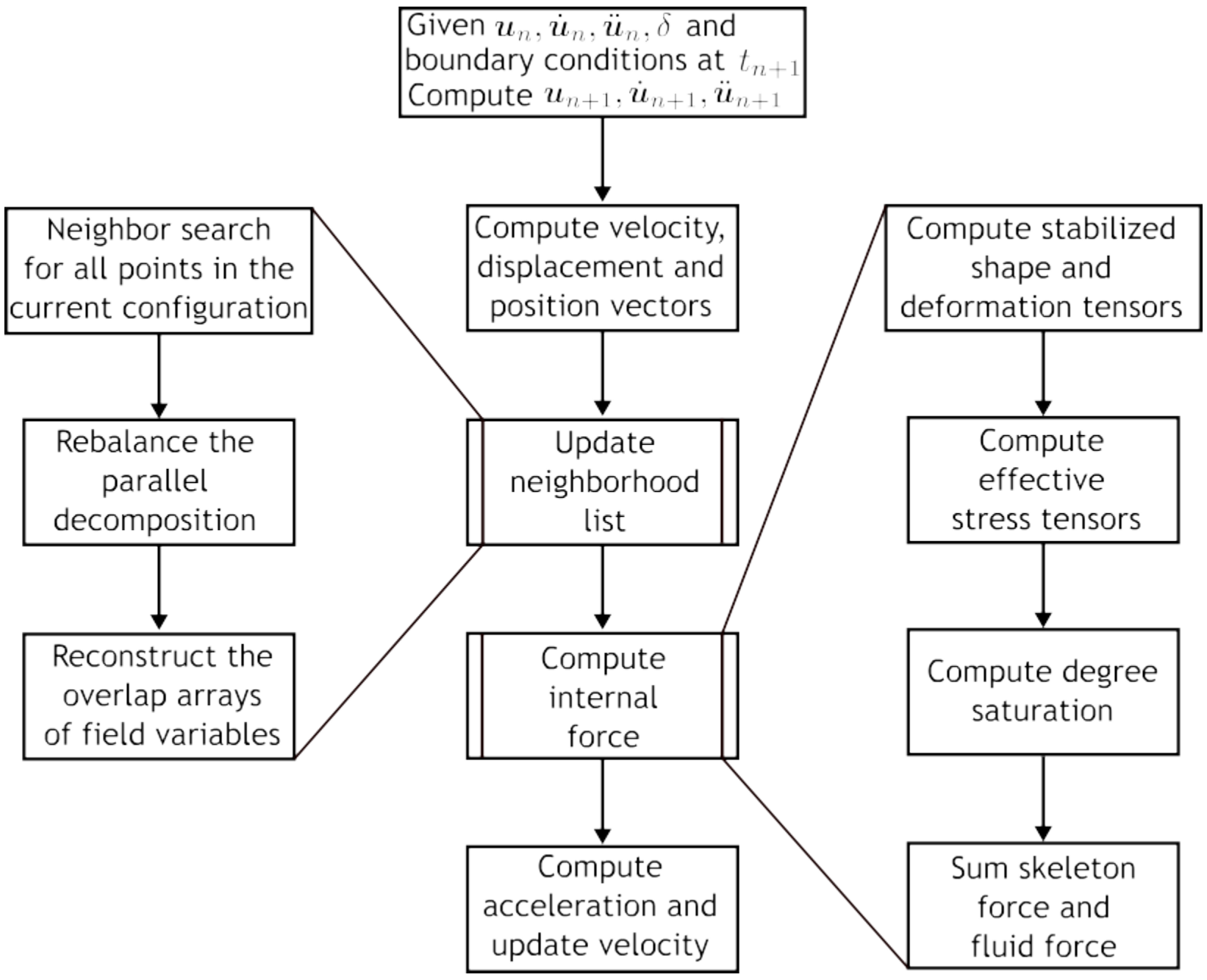

Figure 3: Global flowchart for the explicit numerical algorithm for the updated Lagrangian periporomechanics.

Let $\mathcal{P}$ denote the number of total material points in the problem domain and $\mathcal{N}$ be the number of material points in the horizon of material point $i$. The spatially discretized equation of motion can be written as

$$\mathscr{A}_{i=1}^{\mathcal{P}}(\mathscr{M}_i \ddot{\mathbf{u}}_i - \overline{\mathscr{T}}_i + \mathscr{T}_{w,i} + \mathscr{F}_i) = 0, \tag{47}$$

where $\mathscr{A}$ is a global linear assembly operator [2, 51], $\mathscr{M}_i$ is the mass matrix at material point $i$, $\overline{\mathscr{T}}_i$ is the vector of effective force, $\mathscr{T}_{w,i}$ is the vector of fluid force, $\mathscr{F}$ is the vector of gravity force. In the current configuration of solid skeleton, the three variables can be written as

$$\mathscr{M}_i = [\rho_s(1-\phi_i) + \rho_w S_{r,i}\phi_i]\,\mathscr{V}_i\mathbf{1}, \tag{48}$$

$$\overline{\mathscr{T}}_i = \sum_{j=1}^{\mathcal{N}_i} \left(\underline{\overline{\mathscr{T}}}_{(ij)} - \underline{\overline{\mathscr{T}}}_{(ji)}\right) \mathscr{V}_j \mathscr{V}_i, \tag{49}$$

$$\mathscr{T}_{w,i} = \sum_{j=1}^{\mathcal{N}_i} \left(S_{r,i}\underline{\mathscr{T}}_{w,(ij)} - S_{r,j}\underline{\mathscr{T}}_{w,(ji)}\right) \mathscr{V}_j \mathscr{V}_i, \tag{50}$$

where $\mathscr{V}_i$ and $\mathscr{V}_j$ are the volumes of material points $i$ and $j$, respectively, in the current configuration.

In (49) and (50), the effective force state and the water force state are written as

$$\overline{\mathscr{T}}_{(ij)} = \varpi_{(ij)}\underline{\omega}_{(ij)}\underline{\boldsymbol{\zeta}}_{(ij)}\mathscr{K}^{-1}_{(ij)}\overline{\boldsymbol{\sigma}}_{(ij)}, \tag{51}$$

$$\overline{\mathscr{T}}_{(ji)} = \varpi_{(ji)}\underline{\omega}_{(ji)}\underline{\boldsymbol{\zeta}}_{(ji)}\mathscr{K}^{-1}_{(ji)}\overline{\boldsymbol{\sigma}}_{(ji)}, \tag{52}$$

$$\underline{\mathscr{T}}_{(w,ij)} = \varpi_{(ij)}\underline{\omega}_{(ij)}\underline{\boldsymbol{\zeta}}_{(ij)}\mathscr{K}^{-1}_{(ij)}\mathbf{1}p_{w,i}, \tag{53}$$

$$\underline{\mathscr{T}}_{(w,ji)} = \varpi_{(ji)}\underline{\omega}_{(ji)}\underline{\boldsymbol{\zeta}}_{(ji)}\mathscr{K}^{-1}_{(ji)}\mathbf{1}p_{w,i}, \tag{54}$$

The weighting factors $\varpi_{(ij)}$ and $\varpi_{(ji)}$ in (51) – (54),

$$\varpi_{(ij)} = \frac{\sum_{k=1}^{\mathcal{N}_{(ij)}} \mathscr{V}_k}{\sum_{l=1}^{\mathcal{N}_i} \mathscr{V}_l}, \tag{55}$$

$$\varpi_{(ji)} = \frac{\sum_{k=1}^{\mathcal{N}_{(ji)}} \mathscr{V}_k}{\sum_{l=1}^{\mathcal{N}_j} \mathscr{V}_l}, \tag{56}$$

where $\mathscr{N}_{(ij)}$ is the number of material points in the sub-horizon $\mathscr{H}_{(ij)}$ and $\mathscr{N}_{(ji)}$ is the number of material points in the sub-horizon $\mathscr{H}_{(ji)}$.

The velocity gradient is written as

$$\boldsymbol{\mathscr{L}}_{(ij)} = \left[\sum_{k=1}^{\mathcal{N}_{(ij)}} \left(\underline{\omega}_{(ij)}\dot{\boldsymbol{\mathscr{Y}}}_{(ik)} \otimes \underline{\boldsymbol{\zeta}}_{(ik)}\right)\mathscr{V}_k\right](\boldsymbol{\mathscr{K}}_{(ij)})^{-1}. \tag{57}$$

The rate of deformation tensor reads

$$\boldsymbol{\mathscr{D}}_{(ij)} = \frac{1}{2}\left[\boldsymbol{\mathscr{L}}_{(ij)} + \boldsymbol{\mathscr{L}}_{(ji)}\right], \tag{58}$$

Then the increment in strain reads

$$\boldsymbol{\varepsilon}_{(ij)} = \Delta t\,\boldsymbol{\mathscr{D}}_{(ij)}. \tag{59}$$

Given (59) a classical constitutive model for unsaturated porous media can be used to compute $\overline{\boldsymbol{\sigma}}_{(ij)}$ as described in Section 3.2.1. Next, we introduce the temporal discretization through an explicit Newmark scheme.

### 3.2. *Integration in time*

The Newmark scheme [51] is adopted to integrate the motion of equation in time. Let $\boldsymbol{u}_n$, $\dot{\boldsymbol{u}}_n$ and $\ddot{\boldsymbol{u}}_n$ be the displacement, velocity, and acceleration vectors at time step $n$. The predictors of displacement and velocity in a general Newmark scheme read

$$\tilde{\dot{\boldsymbol{u}}}_{n+1} = \dot{\boldsymbol{u}}_n + (1-\beta_1)\Delta\ddot{\boldsymbol{u}}_n, \tag{60}$$

$$\tilde{\boldsymbol{u}}_{n+1} = \boldsymbol{u}_n + \Delta t\dot{\boldsymbol{u}}_n + \frac{\Delta t^2}{2}(1-2\beta_2)\,\ddot{\boldsymbol{u}}_n, \tag{61}$$

where $\beta_2$ *and* $\beta_1$ are numerical integration parameters. Given (60) and (61), the acceleration $\ddot{\boldsymbol{u}}_{n+1}$, is determined by the recursion relation,

$$\ddot{\boldsymbol{u}}_{n+1} = \boldsymbol{\mathscr{M}}^{-1}_{n+1}(\boldsymbol{\mathcal{F}}_{n+1} - \widetilde{\widetilde{\boldsymbol{\mathscr{T}}}}_{n+1} + \widetilde{\boldsymbol{\mathscr{T}}}_{w,n+1}), \tag{62}$$

where $\widetilde{\widetilde{\boldsymbol{\mathscr{T}}}}_{n+1}$ and $\widetilde{\boldsymbol{\mathscr{T}}}_{w,n+1}$ are determined from (61) and the local constitutive model. From (62), the displacement and velocity at time step $n+1$ can be updated as

$$\dot{\boldsymbol{u}}_{n+1} = \tilde{\dot{\boldsymbol{u}}}_{n+1} + \beta_1\Delta t\ddot{\boldsymbol{u}}_{n+1}, \tag{63}$$

$$\boldsymbol{u}_{n+1} = \tilde{\boldsymbol{u}}_{n+1} + \beta_2\Delta t^2\ddot{\boldsymbol{u}}_{n+1}. \tag{64}$$

In this study, we adopt the explicit central difference solution scheme [51] in which $\beta_1 = 0.5$ and $\beta_2 = 0$. We note that the explicit method is efficient and robust to model dynamic problems

**Algorithm 1** Given $\boldsymbol{u}$ construct effective force vector $\overline{\mathscr{T}}$ and pressure force vector $\mathscr{T}_w$

Execute neighbor search: $\forall j \in \mathscr{B}$, if $|\underline{\boldsymbol{\zeta}}_{(ij)}| \leq \delta_i$ add $j$ to the neighbor list of $i$, the set $\mathscr{H}_i$
**for** all points $i$ **do**
    **for** all neighboring points $j$ **do**
        Search for the sub-horizon neighbors: $\forall k \in \mathscr{H}_i$, if $|\underline{\boldsymbol{\zeta}}_{jk}| \leq \delta_j$, then $k \in \mathscr{H}_{(ij)}$
        **for** all sub-horizon neighbors $k$ **do**
            Compute contribution of bond $ik$ to the shape tensor $\mathscr{K}_{(ij)}$ using (23)
        **end for**
        Compute the sub-horizon based deformation gradient $\mathscr{F}_{(ij)}$ using (24)
    **end for**
    Compute deformation gradient $\mathscr{F}_i$ and its determinant $\mathscr{J}_i$ using (17)
**end for**
**for** all points $i$ **do**
    **for** all neighboring points $j$ **do**
        **for** all sub-horizon neighbors $k$ **do**
            Compute the contribution of bond $ik$ to the velocity gradient $\mathscr{L}_{(ij)}$ using (25)
        **end for**
        Compute the rate of deformation $\mathscr{D}_{(ij)}$ using (26)
        Compute the strain increment using (59)
        Compute Cauchy stress $\overline{\boldsymbol{\sigma}}_{(ij)}$ at $n+1$
    **end for**
    Compute the degree saturation $S_r$ using (15)
**end for**
**for** all points $i$ **do**
    **for** all neighboring points $j$ **do**
        Compute $\underline{\overline{\boldsymbol{T}}}_{(ij)}$ using (51)
        Compute $\underline{\boldsymbol{T}}_{w,ij}$ using (53)
    **end for**
**end for**
**for** all points $i$ **do**
    Compute $\overline{\mathscr{T}}_i$ using (49)
    Compute $\mathscr{T}_{w,i}$ using (50)
**end for**

with extreme large deformation [20]. Figure 4 plots a flowchart of the algorithm and Algorithm 2 summarizes the central difference time integration scheme.

For the numerical stability of the explicit algorithm, the critical time step is determined through a simple method originally formulated for the bond-based peridynamics for solids (e.g., [20])

$$\Delta t_c = \sqrt{\frac{2(1-\phi)\rho_s}{\sum_j^{\mathcal{N}_i} \mathscr{V}_j \mathcal{C} |\boldsymbol{\zeta}_{(ij)}|}}, \tag{65}$$

where $\varphi$ is the porosity and $\rho_s$ is the intrinsic density of solid skeleton as defined previously, a nd $\mathcal{C}_{ij}$ is the micro-mechanical elastic modulus for a bond $ij$ [12, 20]. We note that (65) can provide a conservative estimate for state-based peridynamics [52]. However, for the extreme large deformation analysis involving visco-plasticity and contact the numerical instability could occur even when (65) is followed. Thus we perform an energy balance check to ensure stability of the algorithm. The internal energy, external energy and kinectic energy of the system at

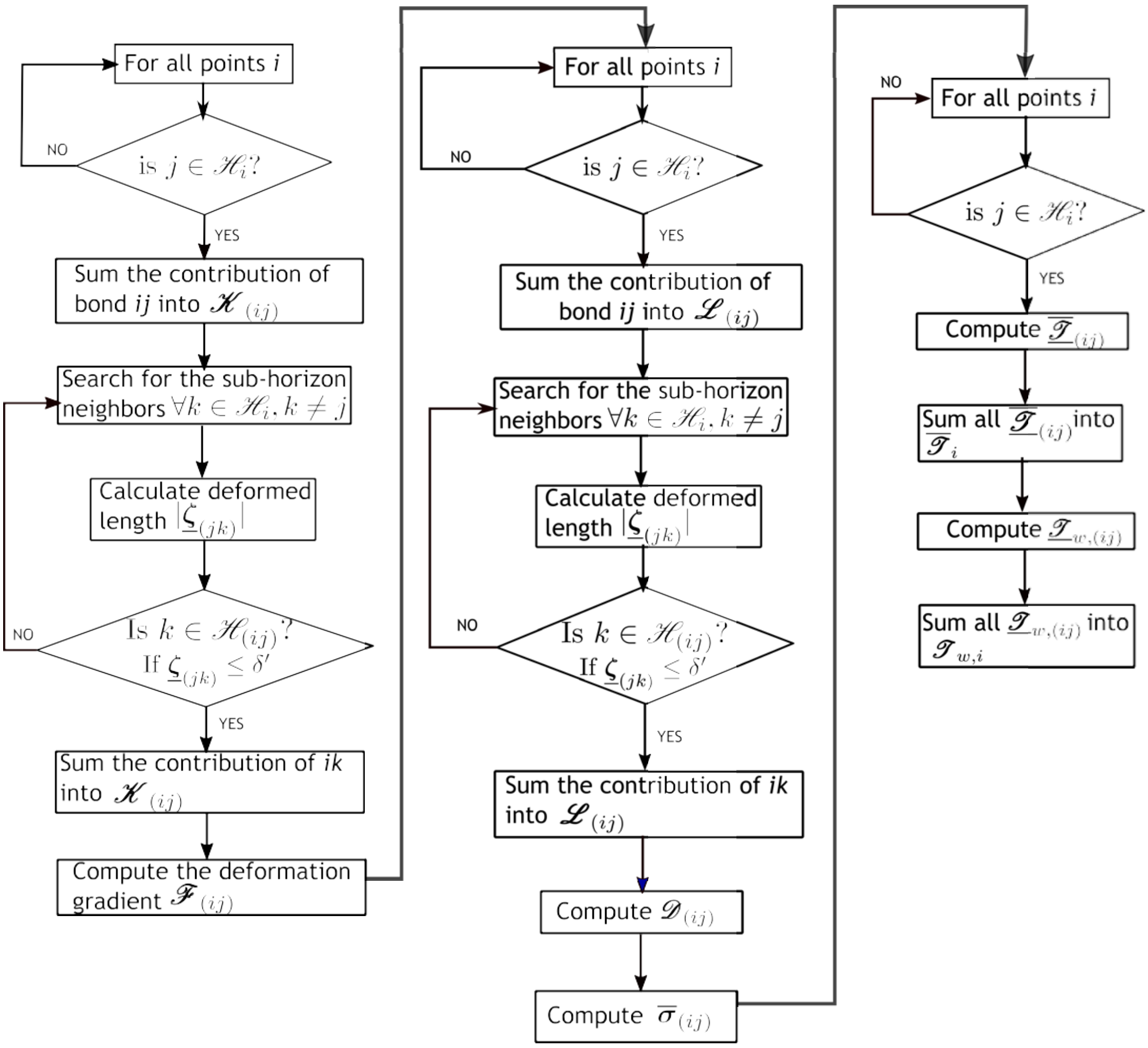

Figure 4: Flowchart for the computation at the material point level for the sub-horizon based updated La-grangian periporomechanics.

time step $n+1$ can be written as

$$\mathscr{W}_{\text{int},n+1} = \mathscr{W}_{\text{int},n} + \frac{\Delta t}{2}\left(\dot{\boldsymbol{u}}_n + \frac{1}{2}\Delta t \ddot{\boldsymbol{u}}_n\right)\left[\left(\overline{\mathscr{T}}_n - \mathscr{T}_{w,n}\right) + \left(\overline{\mathscr{T}}_{n+1} - \mathscr{T}_{w,n+1}\right)\right], \tag{66}$$

$$\mathscr{W}_{\text{ext},n+1} = \mathscr{W}_{\text{ext},n} + \frac{\Delta t}{2}\left(\dot{\boldsymbol{u}}_n + \frac{1}{2}\Delta t \ddot{\boldsymbol{u}}_n\right)\left(\mathscr{F}_n + \mathscr{F}_{n+1}\right), \tag{67}$$

$$\mathscr{W}_{\text{kin},n+1} = \frac{1}{2}\dot{\boldsymbol{u}}_{n+1}\mathscr{M}_{n+1}\dot{\boldsymbol{u}}_{n+1}. \tag{68}$$

Then energy conservation requires that

$$\left|\mathscr{W}_{\text{int},n+1} + \mathscr{W}_{\text{kin},n+1} - \mathscr{W}_{\text{ext},n+1}\right| \leq \bar{\varepsilon}\, \max(\mathscr{W}_{\text{kin},n+1}, \mathscr{W}_{\text{int},n+1}, \mathscr{W}_{\text{ext},n+1}), \tag{69}$$

where $\varepsilon$ is a small tolerance on the order of $10^{-2}$ [53] .

For self-completeness, in what follows we summarize the numerical integration algorithm for the local constitutive model.

### 3.2.1. *Integration of the local constitutive model*

The backward Euler integration scheme is adopted to numerically implement the adopted viscoplaticity model for unsaturated soil in the updated Lagrangian periporomechanics model (e.g., [1]). Given $\Delta\boldsymbol{\varepsilon}_{n+1} = \boldsymbol{\varepsilon}_{n+1} - \boldsymbol{\varepsilon}_n$, the trial effective stress tensor can be computed

**Algorithm 2** Explicit Newmark time integration scheme

Set initial conditions $\dot{\boldsymbol{u}}_0$, $\boldsymbol{u} = 0$, $t_n = 0$, $\overline{\boldsymbol{\sigma}}_0$, $\boldsymbol{p}_{w0}$ and compute $\Delta t = \Delta t_{crit}$ using (65)
Evaluate initial state using Algorithm 1
Compute initial acceleration $\ddot{\boldsymbol{u}}_0 = \boldsymbol{\mathscr{M}}_0^{-1}(-\overline{\boldsymbol{\mathscr{T}}}_0 + \boldsymbol{\mathscr{T}}_{w,0})$
**while** $t_n < t_{\text{final}}$ **do**
    Update time $t_{n+1} = t_n + \Delta t$
    Compute the predictor $\tilde{\dot{\boldsymbol{u}}}_{n+1}$ using (60)
    Apply boundary conditions
    Compute the displacement $\boldsymbol{u}_{n+1}$ using (64)
    Compute internal force using Algorithm in Section 3.2.1
    Compute the acceleration $\ddot{\boldsymbol{u}}_{n+1}$ using (62)
    Compute the velocity $\dot{\boldsymbol{u}}_{n+1}$ using (63)
    Compute global kinetic energy $\mathscr{W}_{\text{kin},n+1}$ internal energy $\mathscr{W}_{\text{int},n+1}$ and external energy $\mathscr{W}_{\text{ext},n+1}$ using (66), (67) and (68)
    Check energy balance
    $n \leftarrow n + 1$
**end while**
**end**

from the elastic model. For the elastic loading case, the trial effective stress tensor is the real effective stress state. For the visco-plastic loading case (i.e., $f > 0$), the effective stress state and apparent pre-consolidation pressure can be solved in the stress invariant $(\bar{p}, q)$ space as follows. At time step $n+1$, $\bar{p}$, $q$, and $p_c$ can be written as

$$\bar{p} = \bar{p}^{tr} - K\frac{f\Delta t}{\eta}\left(\frac{\partial f}{\partial \bar{p}}\right), \tag{70}$$

$$q = q^{tr} - 3\mu_s\frac{f\Delta t}{\eta}\left(\frac{\partial f}{\partial q}\right), \tag{71}$$

$$\bar{p}_c = \bar{p}_{c,n}\exp\left[\frac{f\Delta t}{(\tilde{\lambda} - \tilde{\kappa})\eta}\left(\frac{\partial f}{\partial \bar{p}}\right)\right]. \tag{72}$$

The values of $\bar{p}, q, \bar{p}_c$ at time step $n+1$ can be determined through Newton's method by defining a residual vector $\boldsymbol{r} = \{r_1, r_2, r_3\}^T$ as

$$r_1 = \bar{p} - \bar{p}^{tr} + K\frac{f\Delta t}{\eta}\left(\frac{\partial f}{\partial \bar{p}}\right), \tag{73}$$

$$r_2 = q - q^{tr} + 3\mu_s\frac{f\Delta t}{\eta}\left(\frac{\partial f}{\partial q}\right), \tag{74}$$

$$r_3 = \bar{p}_c - \bar{p}_{c,n}\exp\left[\frac{f\Delta t}{(\tilde{\lambda} - \tilde{\kappa})\eta}\left(\frac{\partial f}{\partial \bar{p}}\right)\right], \tag{75}$$

where $\bar{p}^{\text{tr}}$ and $q^{\text{tr}}$ are the trial values by freezing the vicoplastic deformation at time step $n+1$ [49]. Let the local unknown vector be $\boldsymbol{x} = \{\bar{p}, q, \bar{p}_c\}^T$, through Newtons' method $\boldsymbol{x}$ can be solved as

$$\boldsymbol{x}^{k+1} = \boldsymbol{x}^k - \boldsymbol{A}^{-1}\boldsymbol{r}^k(\boldsymbol{x}). \tag{76}$$

where $k$ is the iteration counter and $\boldsymbol{A}$ is the local tangent operator that reads

$$\boldsymbol{A} = \begin{bmatrix} A_{11} & A_{12} & A_{13} \\ A_{21} & A_{22} & A_{23} \\ A_{31} & A_{32} & A_{33} \end{bmatrix}. \tag{77}$$

The individual elements of $\boldsymbol{A}$ can be written as

$$A_{11} = 1 + K\frac{\Delta t}{\eta}\left(\frac{\partial f}{\partial \bar{p}_c}\frac{\partial f}{\partial \bar{p}} + f\frac{\partial^2 f}{\partial \bar{p}_c \partial \bar{p}}\right)\frac{\partial \bar{p}_c}{\partial \bar{p}}, \tag{78}$$

$$A_{12} = K\frac{\Delta t}{\eta}\left(\frac{\partial f}{\partial q}\frac{\partial f}{\partial \bar{p}}\right), \tag{79}$$

$$A_{13} = K\frac{\Delta t}{\eta}\left(\frac{\partial f}{\partial \bar{p}_c}\frac{\partial f}{\partial \bar{p}} + f\frac{\partial^2 f}{\partial \bar{p}_c \partial \bar{p}}\right)\frac{\partial \bar{p}_c}{\partial p_c}, \tag{80}$$

$$A_{21} = 3\mu_s\frac{\Delta t}{\eta}\left(\frac{\partial f}{\partial \bar{p}_c}\frac{\partial f}{\partial q}\right)\frac{\partial \bar{p}_c}{\partial \bar{p}}, \tag{81}$$

$$A_{22} = 1 + 3\mu_s\frac{\Delta t}{\eta}\left[f\frac{\partial^2 f}{\partial q^2} + \left(\frac{\partial f}{\partial q}\right)^2\right], \tag{82}$$

$$A_{23} = 3\mu_s\frac{\Delta t}{\eta}\left(\frac{\partial f}{\partial \bar{p}_c}\frac{\partial f}{\partial q}\right)\frac{\partial \bar{p}_c}{\partial p_c}, \tag{83}$$

$$A_{31} = -\widetilde{p}_c\frac{\Delta t}{(\tilde{\lambda} - \tilde{\kappa})\eta}\left(\frac{\partial f}{\partial \bar{p}_c}\frac{\partial f}{\partial \bar{p}} + f\frac{\partial^2 f}{\partial \bar{p}_c \partial \bar{p}}\right)\frac{\partial \bar{p}_c}{\partial \bar{p}}, \tag{84}$$

$$A_{32} = -\widetilde{p}_c\frac{\Delta t}{(\tilde{\lambda} - \tilde{\kappa})\eta}\left(\frac{\partial f}{\partial q}\frac{\partial f}{\partial \bar{p}}\right), \tag{85}$$

$$A_{33} = -\widetilde{p}_c\frac{\Delta t}{(\tilde{\lambda} - \tilde{\kappa})\eta}\left(\frac{\partial f}{\partial \bar{p}_c}\frac{\partial f}{\partial \bar{p}} + f\frac{\partial^2 f}{\partial \bar{p}_c \partial \bar{p}}\right)\frac{\partial \bar{p}_c}{\partial p_c}, \tag{86}$$

where

$$\frac{\partial f}{\partial \bar{p}_c} = -\bar{p}, \tag{87}$$

$$\frac{\partial^2 f}{\partial \bar{p}_c \partial \bar{p}} = -1, \tag{88}$$

$$\frac{\partial \bar{p}_c}{\partial p_c} = b_2 \exp(b_1)(-p_c)^{b_2 - 1}, \tag{89}$$

$$\frac{\partial \bar{p}_c}{\partial \bar{p}} = \frac{b_2 \Delta t \bar{p}_c}{\eta(\tilde{\lambda} - \tilde{\kappa})}\left[\left(\frac{\partial f}{\partial \bar{p}}\right)^2 + \frac{\partial^2 f}{\partial \bar{p}^2}\right]. \tag{90}$$

## 4. Numerical simulations

### 4.1. Uniaxial compression test

This example deals with the stability analysis of the sub-horizon based updated Lagrangian periporomechanics model. We simulate the uniaxial compression of a rectangular specimen with zero matric suction under quasi-static and dynamic conditions. The problem geometry and loading protocol are presented in Figure 5. The problem is discretized into 12,000 uniform material points. The distance between two adjacent material point centers is $\Delta$ = 0.005 m. An isotropic elastic correspondence constitutive model is utilized for the solid skeleton. The material parameters adopted are bulk modulus $K$ = 25 MPa, shear modulus $\mu_s$ = 15 MPa, solid skeleton density $\rho_s$ = 2200 kg/m$^3$ and initial porosity $\varphi$ = 0.2. The simulation is repeated using two horizons, i.e., $\delta_1$ = 0.01 m and $\delta_2$ = 0.015 m. The ratios of $\delta/\Delta$ are $m$ = 2 and $m$ = 3, respectively.

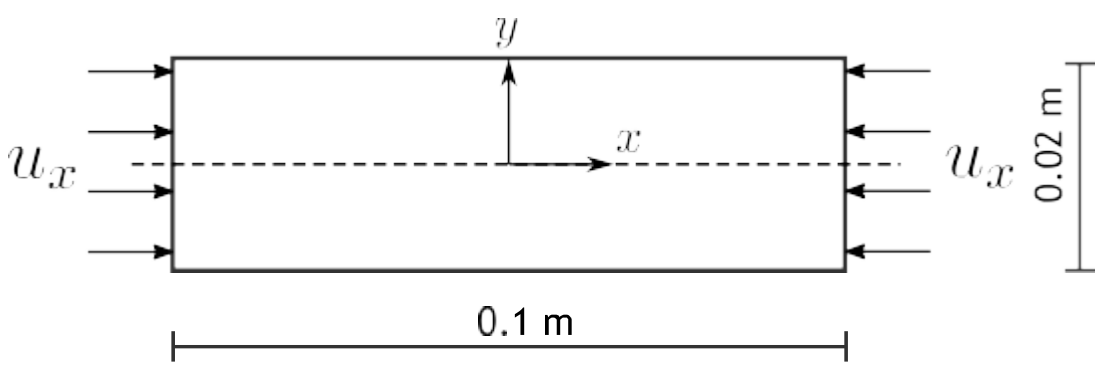

Figure 5: Problem setup for the uniaxial compression test.

The total applied displacement is $u_x$ = 0.5×10$^{-3}$ m. The total loading time is $t$ = 0.1 s. The time increment for the quasi-static simulation is $\Delta t_1$ = 0.001 s and the time increment for the dynamic loading case is $\Delta t_2$ = 3 ×10$^{-6}$ s that meets the critical time step criterion. The numerical results are presented in Figures 6 - 11. For clarity, in what follows the stabilized correspondence constitutive model means that local constitutive models for porous media is implemented using the sub-horizon based correspondence principle formulated in this study. The standard correspondence constitutive model means that the local constitutive model for porous media is implemented using the original multiphase correspondence principle [9**?** ].

Figure 6 presents the contours of $\overline{\sigma}_x$ from the simulations through the standard correspondence constitutive model with two values of $m$. For comparison, Figure 7 plots the contours of $\overline{\sigma}_x$ from the simulations using the stabilized correspondence constitutive model. The con- tours of $\overline{\sigma}_x$ in Figure 6 show noticeable oscillations. Figure 7 shows that the oscillations have disappeared in the results with the stabilized correspondence constitutive model.

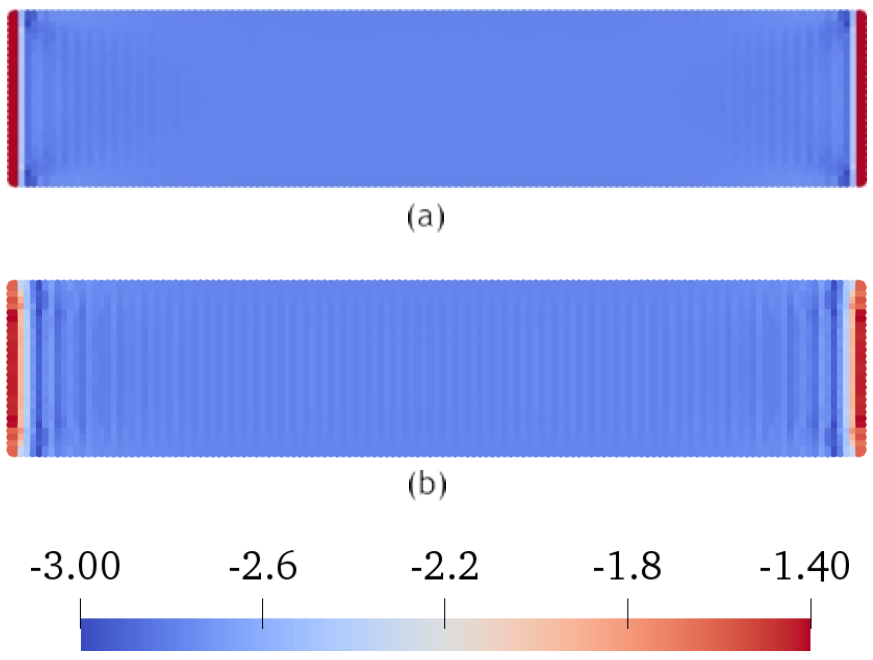

Figure 6: Contours of $\overline{\sigma}_x$ (× 100 kPa) from the quasi-static simulations using the original correspondence elastic constitutive model with (a) $m$ = 2 and (b) $m$ = 3 at $u_x$ = 0.5 × 10$^{-3}$ m.

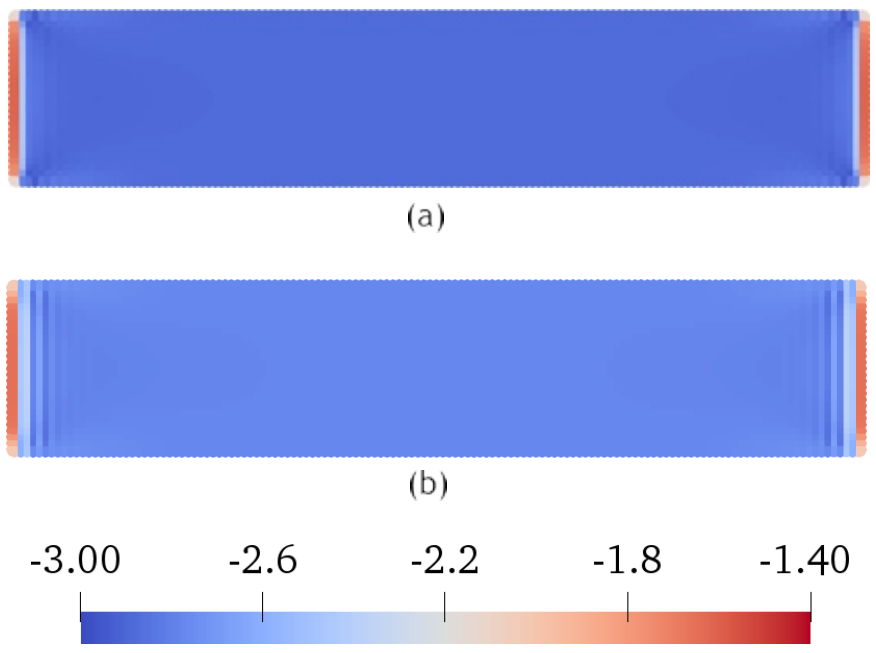

Figure 7: Contours of $\overline{\sigma}_x$ (× 100 kPa) from the quasi-static simulations using the stabilized correspondence constitutive model with (a) $m$ = 2 and (b) $m$ = 3 at $u_x$ = 0.5 × 10$^{-3}$ m.

Figure 8 compares the variations of $\overline{\sigma}_x$ along the $x$ axis of the specimen from the simulations with stabilized and standard correspondence material models. The results show that the higher value of $m$ produces larger oscillations in $\overline{\sigma}_x$ from the simulations using the standard correspondence constitutive model. For the simulations using the stabilized correspondence constitutive model the oscillations have been eliminated for both values of $m$ ratios.

Figures 9 and 10 present the contours of $\overline{\sigma}_x$ from the dynamic simulations with two values of $m$ using the standard and stabilized correspondence constitutive models, respectively. Figure 11 compares the variations of $\overline{\sigma}_x$ along the $x$ axis of the specimen. It is found that the numerical instabilities in the simulations with the standard correspondence constitutive models are more noticeable for the dynamic loading case than the static simulations. Overall, the results in both figures show that the simulations with the stabilized correspondence constitutive model eliminates the oscillations in $\overline{\sigma}_x$.

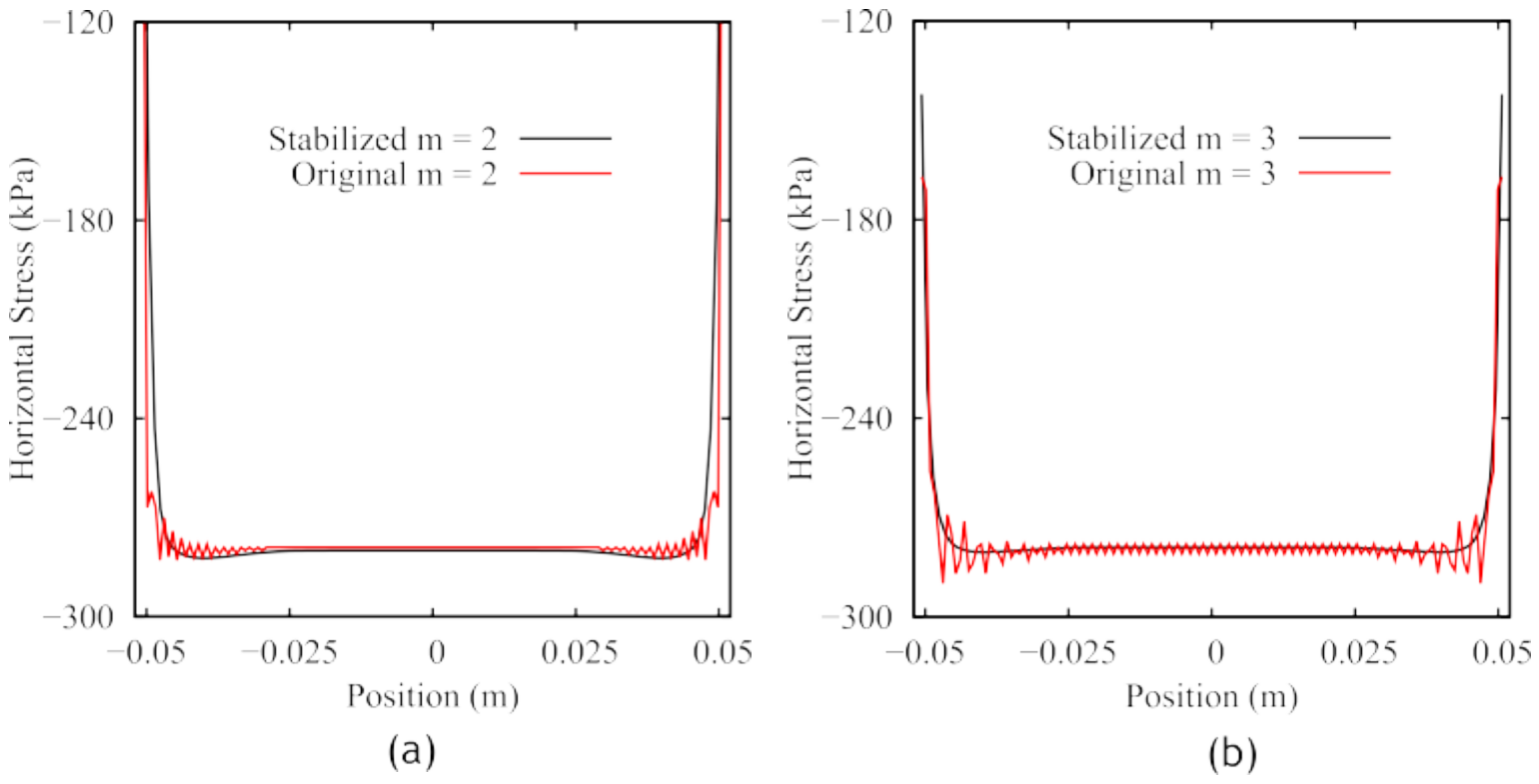

Figure 8: Comparison of $\overline{\sigma}_x$ along the $x$ axis of the specimen from the static loading case with (a) $m$=2 and (b) $m = 3$.

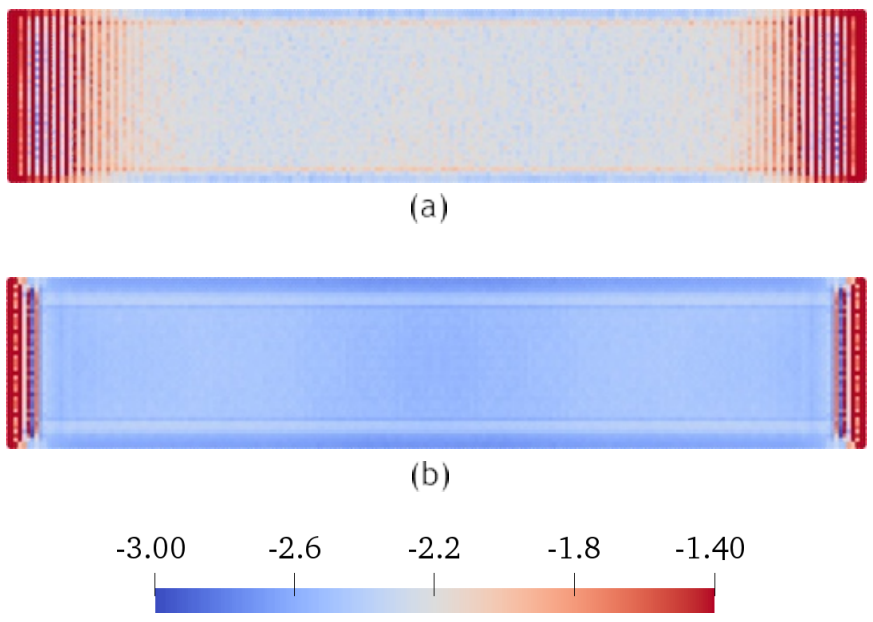

Figure 9: Contours of $\overline{\sigma}_x$ ($\times$ 100 kPa) from the dynamic loading case with (a) $m$ = 2 and (b) $m$ = 3 using the original correspondence constitutive model.

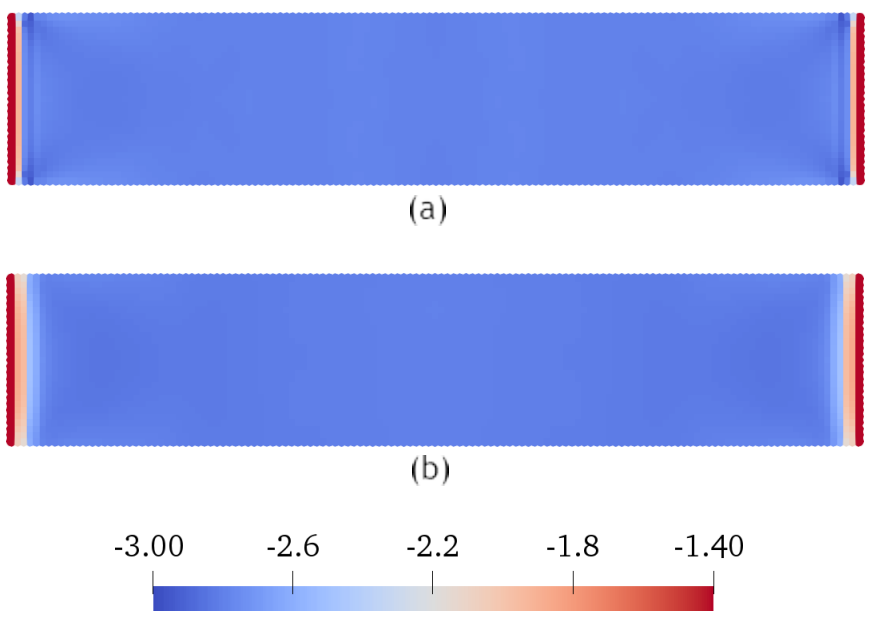

Figure 10: Contours of $\overline{\sigma}_x$ ($\times$ 100 kPa) from the dynamic loading case with (a) $m$ = 2 and (b) $m$ = 3 using the stabilized correspondence constitutive model.

### *4.2. Unsaturated soil column collapse in two dimensions*

In this example, through the proposed sub-horizon based updated Lagrangian periporomechanics we simulate unsaturated soil column collapse by gravity loading under drained condi- tions (i.e., one-way coupled) in two dimensions. We first simulate the collapse of a dry column to demonstrate the ability of the formulation to model extreme large plastic deformation in porous media. We then investigate the influence of the aspect ratio, initial matric suction and sub-grade roughness on the characteristics of collapse. All the simulations are conducted using 64 CPU (central processing unit) cores with a total 512 GB (gigabyte) of dedicated memory. Next, we introduce the problem set up, input material parameters, initial conditions, and the contact model for the soil column and the rigid substrate.

The problem geometry is depicted in Figure 12. The solid skeleton is modeled using the visco-plastic constitutive model introduced in Section 3.2.1. The rigid substrate is modeled using an isotropic elastic model. For the solid skeleton the material parameters are: $K = 25$ MPa, $\mu_s$ = 15 MPa, $\rho_s$ = 2200 kg/m$^3$, $\rho_w$ = 1000 kg/m$^3$, initial porosity $\varphi$ = 0.2 $M$ =

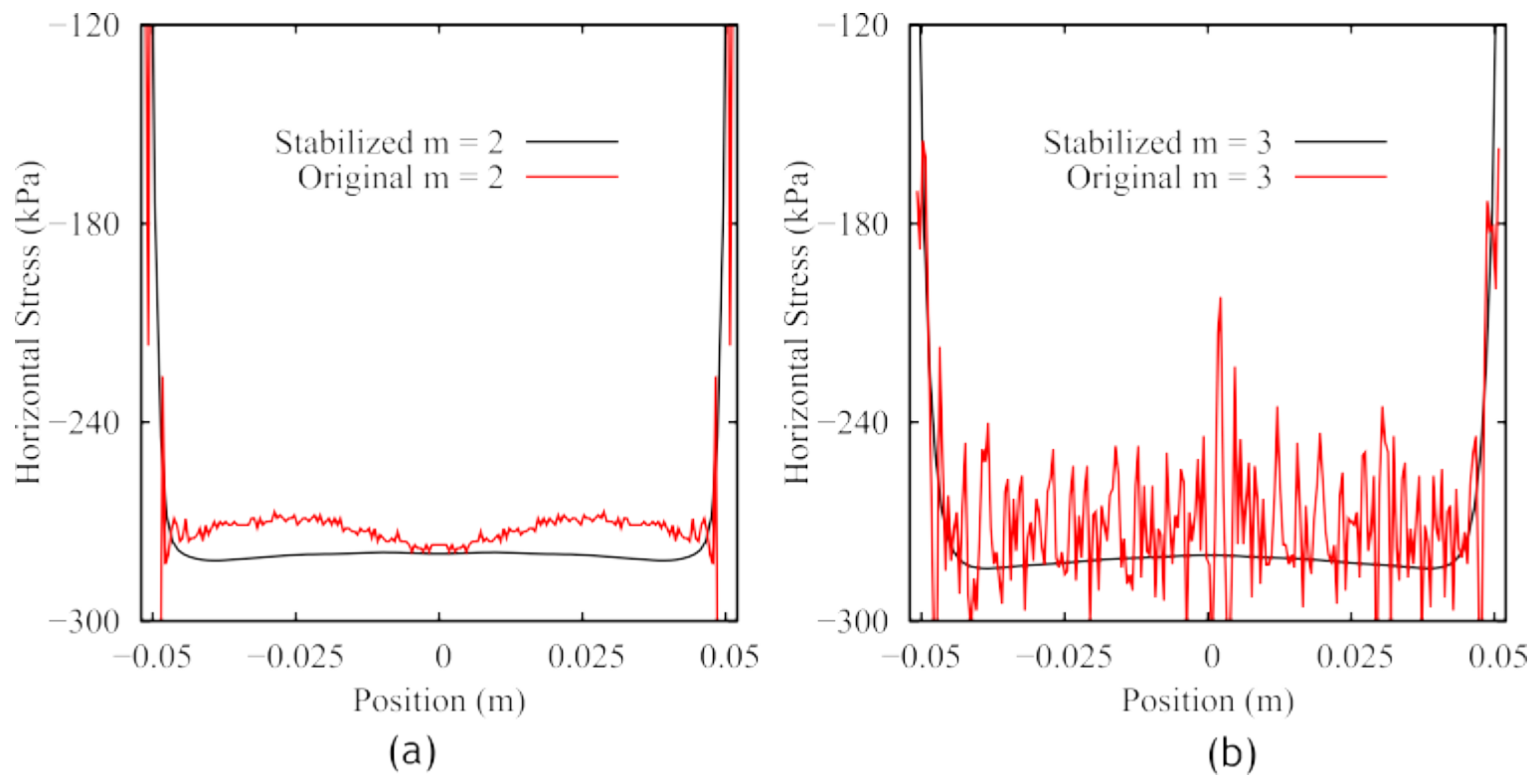

Figure 11: Comparison of $\overline{\sigma}_x$ along the $x$ axis of the specimen from the dynamic loading case with (a) $m$=2 and (b) $m$ = 3.

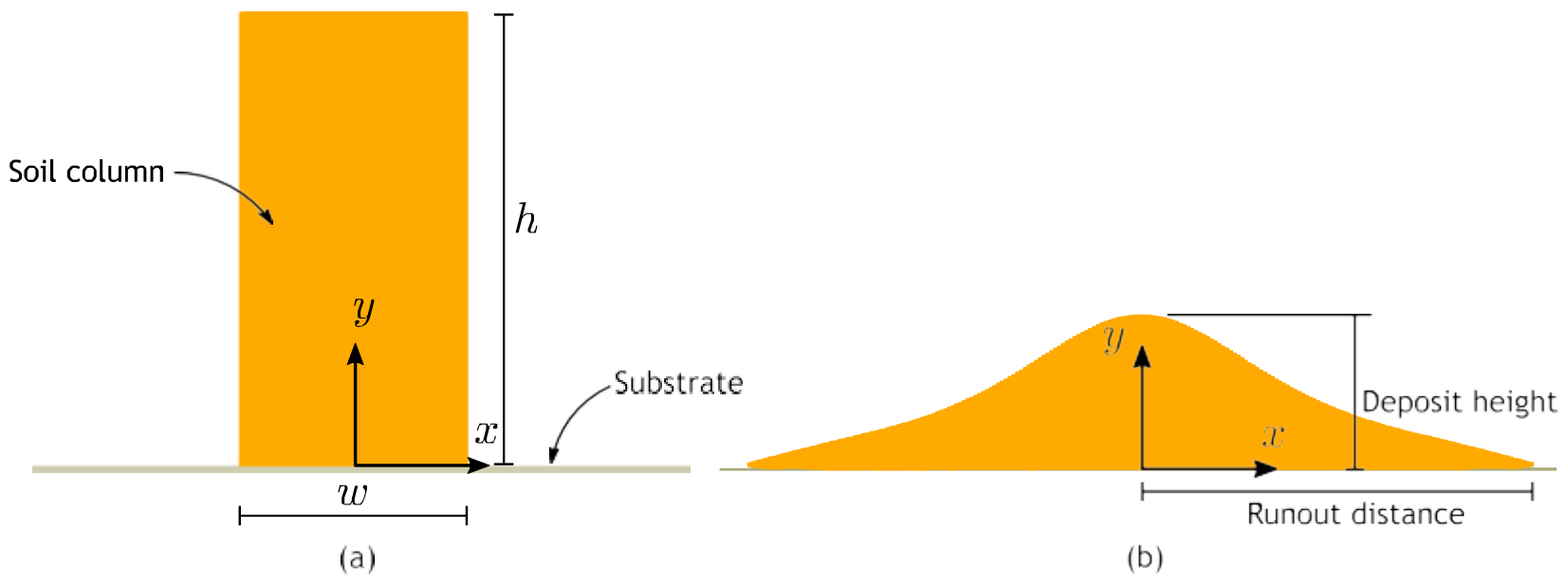

Figure 12: Problem setup for unsaturated soil column collapse in two-dimensions (a) initial configuration, (b) final configuration.

1.1, $\widetilde{\lambda}$ = 0.12, $\widetilde{\kappa}$ = 0.04, $\eta$ = 1000 Pa$^3$⁄s and an over-consolidation ratio of 1.25 (i.e., slightly overconsolidated soil). The parameters for the soil-water retention curve are: $a_1$ = 0.038, $a_2$ = 3.49 and $n$ = 1.25. The initial geostatic stress in the soil is prescribed through a quasi-static loading step. Subsequently, the lateral constraints on the soil column are relaxed to allow it to collapse onto the rigid substrate.

A contact model for the interface of the soil and the substrate is needed to model the spreading of soil over the substrate. In this study we adopt the short range force model in [20? ]. In this contact model, contact interactions are modeled using spring-like repulsive forces acting along the normal to the substrate surface. The contact forces act along virtual bonds that carry only contact force (i.e., no material interaction). The repulsive forces act between pairs of material points within a cut-off distance of each other, $\delta_c$. The repulsive contact force and frictional contact force [54] are defined as

$$\underline{\boldsymbol{T}}_c = -C_s\left(\delta_c - |\underline{\mathscr{Y}}|\right)\frac{\underline{\mathscr{Y}}}{|\underline{\mathscr{Y}}|}, \tag{91}$$

$$\underline{\boldsymbol{T}}_f = -\mu_f \ \mathrm{sign}\left(\frac{\partial}{\partial t}|\underline{\mathscr{Y}}|\right)\underline{\boldsymbol{T}}_c, \tag{92}$$

where $C_s$ is the contact stiffness and $\mu_f$ is the friction coefficient. Given (91) and (92), the equation of motion with the contact model is written as

$$\rho\ddot{\boldsymbol{u}} = \int_{\mathscr{H}_s}\left(\underline{\boldsymbol{T}} - S_r\underline{\boldsymbol{T}}_w\right) - \left(\underline{\boldsymbol{T}} - S_r'\underline{\boldsymbol{T}}_w\right)\,\mathrm{d}\mathscr{V}' + \int_{\mathscr{H}_c}\left[\left(\underline{\boldsymbol{T}}_c + \underline{\boldsymbol{T}}_f\right) - \left(\underline{\boldsymbol{T}}_c' + \underline{\boldsymbol{T}}_f'\right)\right]\,\mathrm{d}\mathscr{V}' + \boldsymbol{b}, \tag{93}$$

where $H_c$ is the contact neighborhood defined by $\delta_c$. Figure 13 depicts the contact interaction between the soil column and the substrate. The initial contact domain is specified in the inputfile of the numerical model. A search algorithm is used to detect contact pairs in a radius $r_c$ around each individual material point in the defined contact domain. The input parameters for the contact model are: $C_s$ = 4 × 10$^6$ N/m$^2$, $\mu_f$ = 0.25, $\delta_c$ = 0.8Δ, and $r_s$ = 3$\delta_c$.

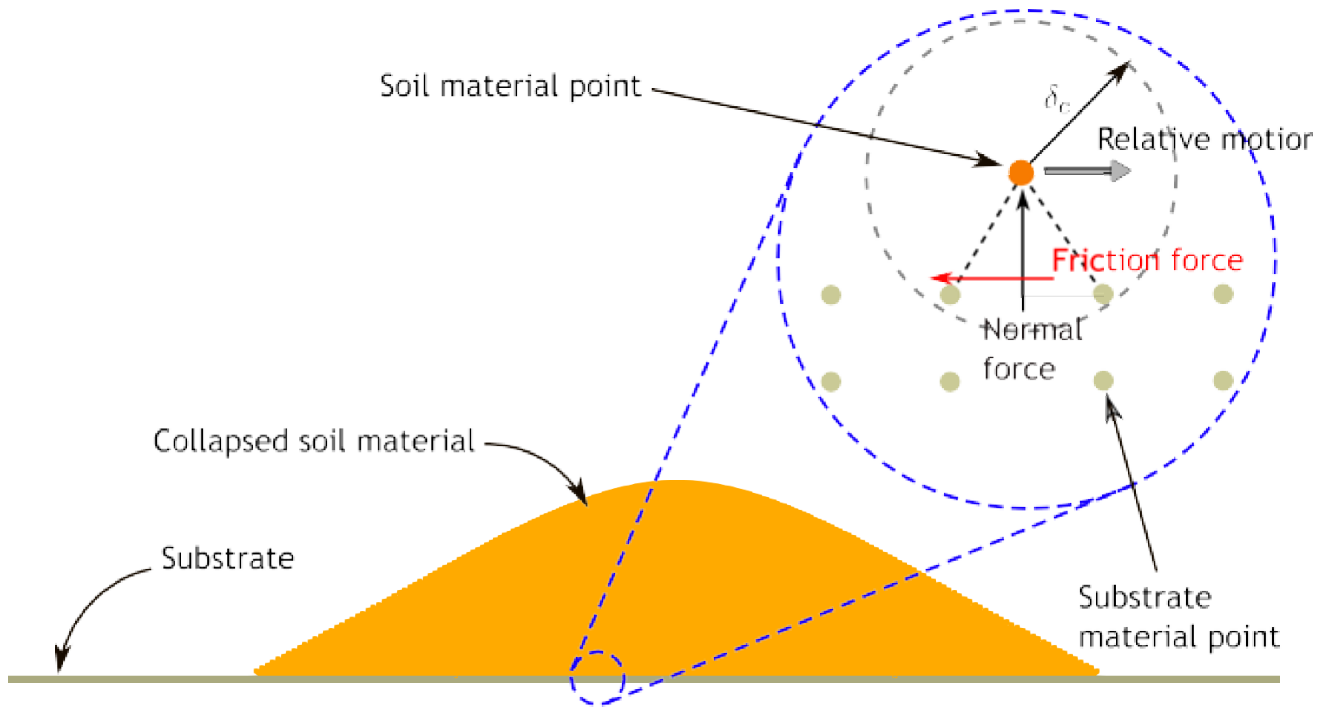

Figure 13: Schematic of the contact model for the soil column and the substrate.

In what follows we study the influence of initial aspect ratio, initial matric suction, and substrate roughness on the soil column collapse.

### 4.2.1. *Influence of initial aspect ratios*

We first study the influence of initial aspect ratios (width over height) on the characteristics (e.g., run-out distance and final deposit height) of the soil column collapse under completely dry conditions. Our numerical results are compared with the experimental data in the literature [36, 37]. We run the simulations with three aspect ratios, $a_1 = 2$, $a_2 = 1$, and $a_3 = 0.5$. All three specimens have the same initial width $w$ 0.1 m. Figure 14 plot the contours of the initial vertical stress in the three specimens.

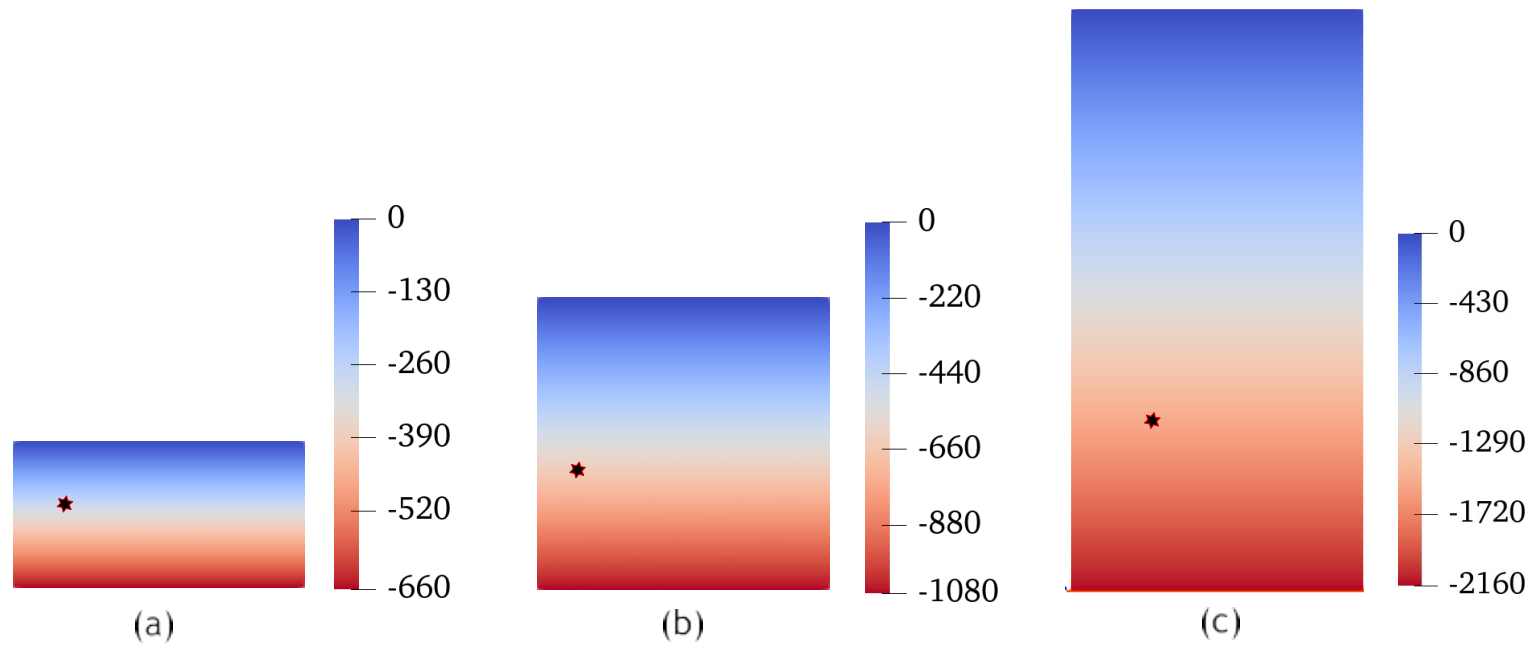

Figure 14: Contours of initial effective vertical stress $\bar{\sigma_y}$ in the specimens with (a) $a_1 = 2$, (b) $a_2 = 1$ and (c) $a_3 = 0.5$.

Figures 15, 16 and 17 plot the snapshots of the contours of $\varepsilon_s$ for the three aspect ratios. The results show that the aspect ratio could affect the soil column collapse processes and final deposit morphologies which are consistent with the experimental results in the literature[36, 37]. As shown in Figures 15 (a) and 16 for the specimens with $a_1 = 2$ and $a_2 = 1$, banded zone of extensive shear develops during the collapse process. For the specimen with $a_3 = 0.5$, the collapse pattern is clearly different from the simulations using larger aspect ratios. The results confirms the existence of two moving layers observed in the experimental testing. Initially, the contour show that the upper half of the column moves directly downward with little horizontal movement, while the base rapidly spreads outward. Subsequently, the upper half of the column moves laterally and creates two lateral moving fronts, the so-called "Mexican-hat" morphology observed in experimental testings of soil column collapse [36, 37]. It is likely that the inertial effects contribute to the change in behavior observed for the simulation with $a_3 = 0.5$. Figures 18, 19 and 20 presents the snapshots of the contour of $\bar{\sigma_y}$ for the three aspect ratios. The results show that the aspect ratio could significantly impact the vertical stress states in the final deposit. Both the contours of equivalent shear strains and vertical stresses demonstrate there could be an underlying and undisturbed region in the soil columnthroughout the collapse process.

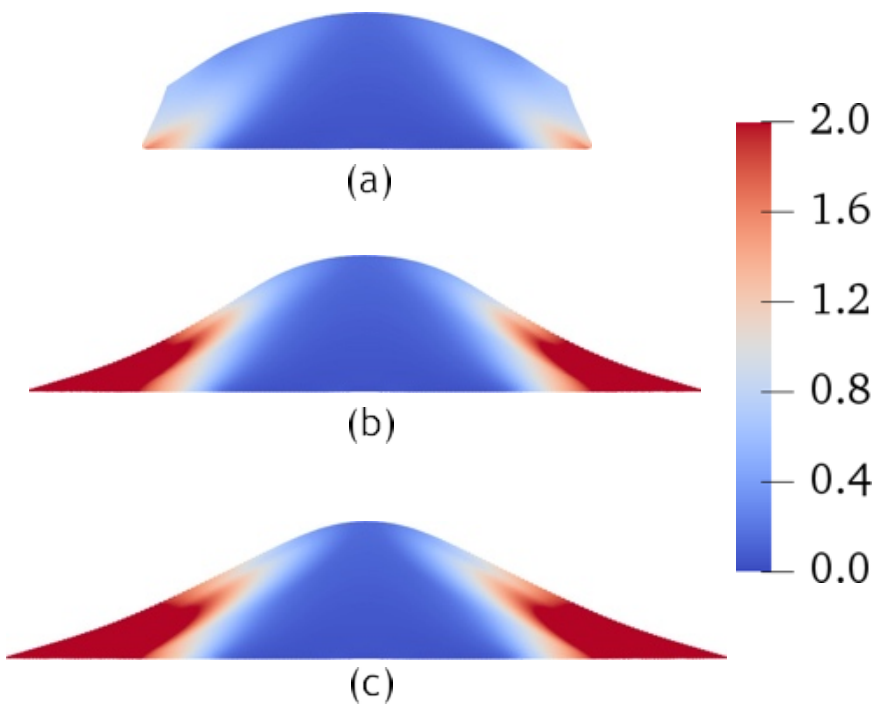

Figure 15: Contours of $\varepsilon_s$ at (a) $t = 0.11$ s, (b) $t = 0.17$ s, (c) $t = 0.25$ s superimposed on the deformed configuration for $a_1 = 2$.

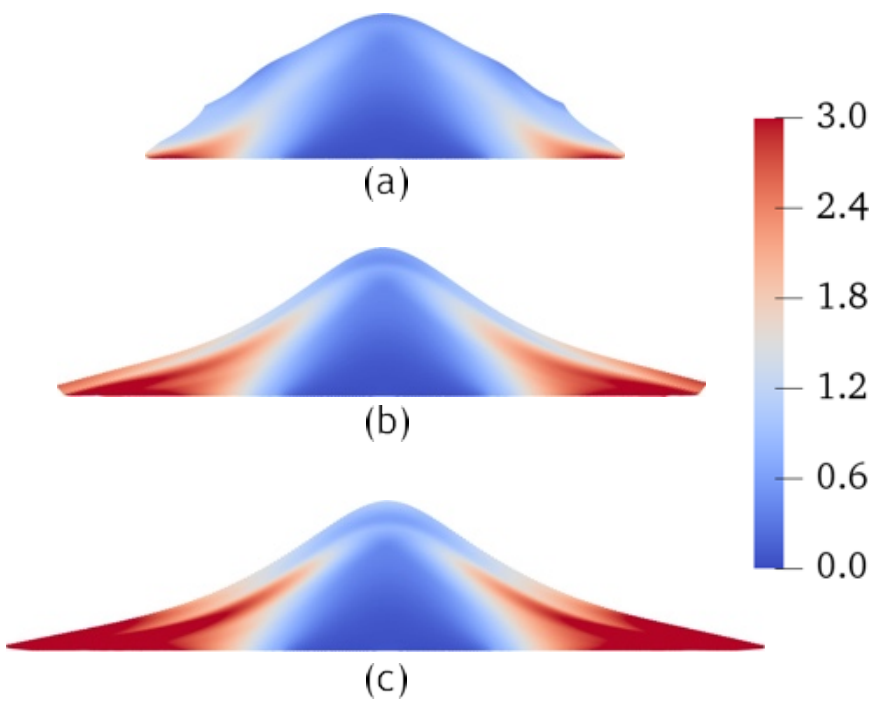

Figure 16: Contours of $\varepsilon_s$ at (a) $t = 0.17$ s, (b) $t = 0.27$ s, (c) $t = 0.4$ s superimposed on the deformed configuration for $a_2 = 1$

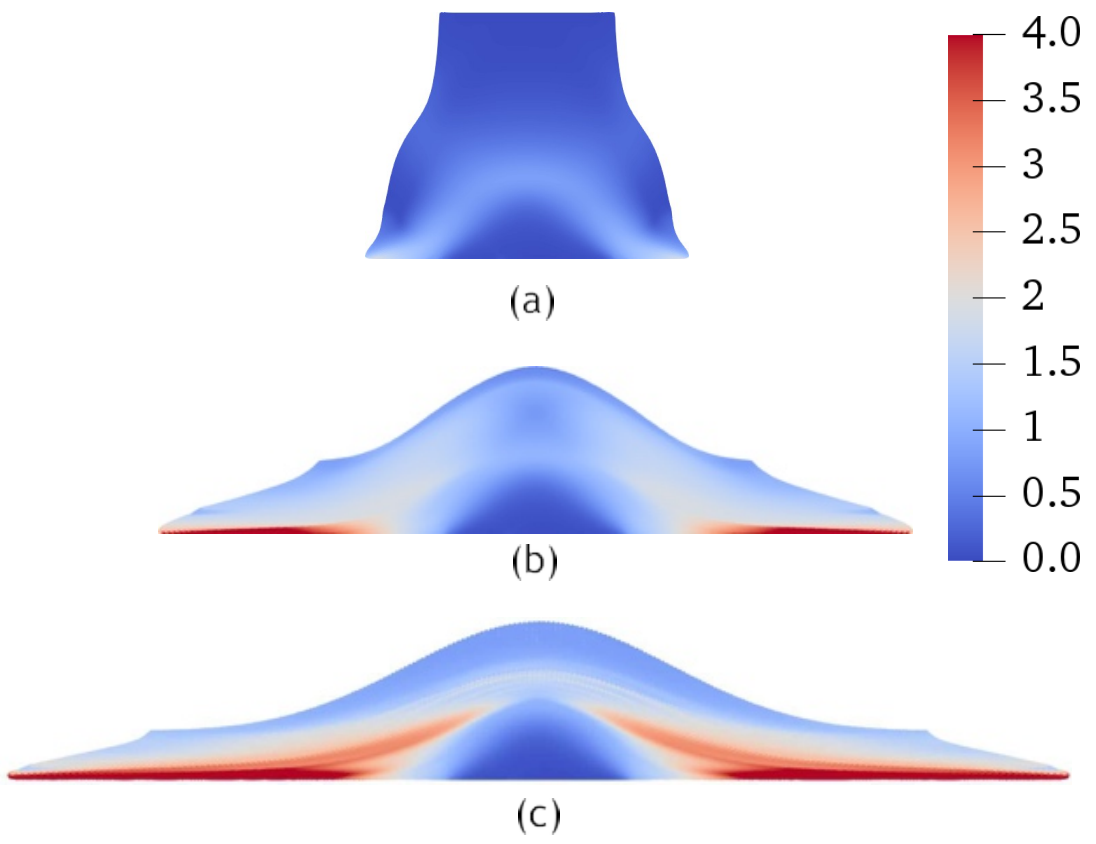

Figure 17: Contours of $\varepsilon_s$ at (a) $t = 0.1$ s, (b) $t = 0.3$ s, (c) $t = 0.5$ s superimposed on the deformed configuration for $a_3 = 2$

Next, we present the contact force along the base in the final deposit configuration for the simulations with three aspect ratios. Figures 21 and 22 plot the normal and frictional components of the contact force acting along the base of the final deposit configuration. The results show that both normal and frictional forces along the interface have oscillations alongthe base of final deposit configurations. The general trend observed from the results in figures21 and 22 is that the normal and shear forces are larger for the specimen with a smaller aspectratio (i.e., taller soil column).

To validate our numerical results, the final deposit height and runout distance are compared against the experiment testing results in the literature. Here the final runout distance is the distance between the lateral front of the final deposit and the center along the substrate. The deposit height is measured from the substrate to the highest point of the final deposit. The final height and runout distance are normalized by the initial width of the soil column. Figure 23 plots the normalized final runout distance from the numerical simulations and the experimental data for three aspect ratios. The numerical results that are consistent with the experimental data demonstrate that the final runout distance decreases with the increase of aspect ratios. Figure 24 compare the normalized finial deposit heights from our numerical simulations and the experimental data for three aspect ratios are presented in the literature. It can be conclude from the results in Figures 23 and 24 that the update Lagrangian periporomechanics formulation can be applied to model soil column collapse under gravity loading that involves extreme large deformation.

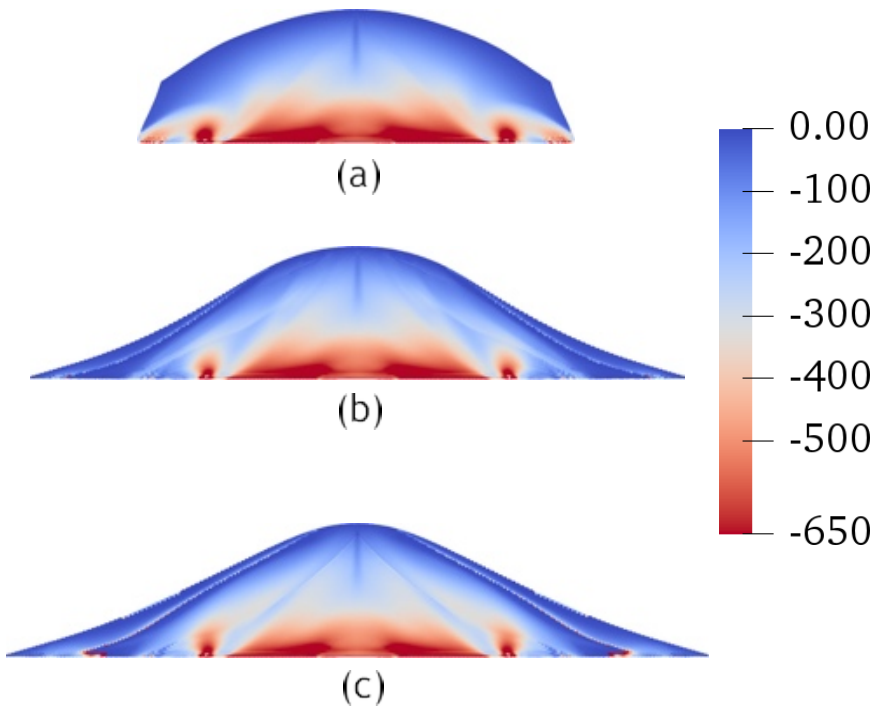

Figure 18: Contours of $\overline{\sigma}_y$ at (a) $t$ = 0.11 s, (b) $t$ = 0.17 s, (c) $t$ = 0.25 s superimposed on the deformed configuration for $a_1 = 2$.

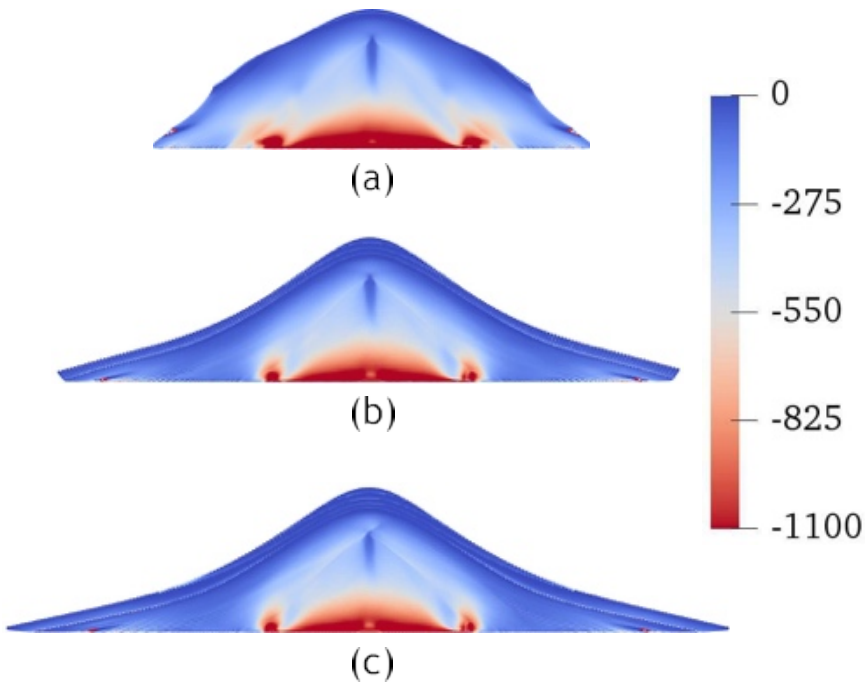

Figure 19: Contours of $\overline{\sigma}_y$ at (a) $t$ = 0.17 s, (b) $t$ = 0.27 s, (c) $t$ = 0.4 s superimposed on the deformed configuration for $a_2 = 1$

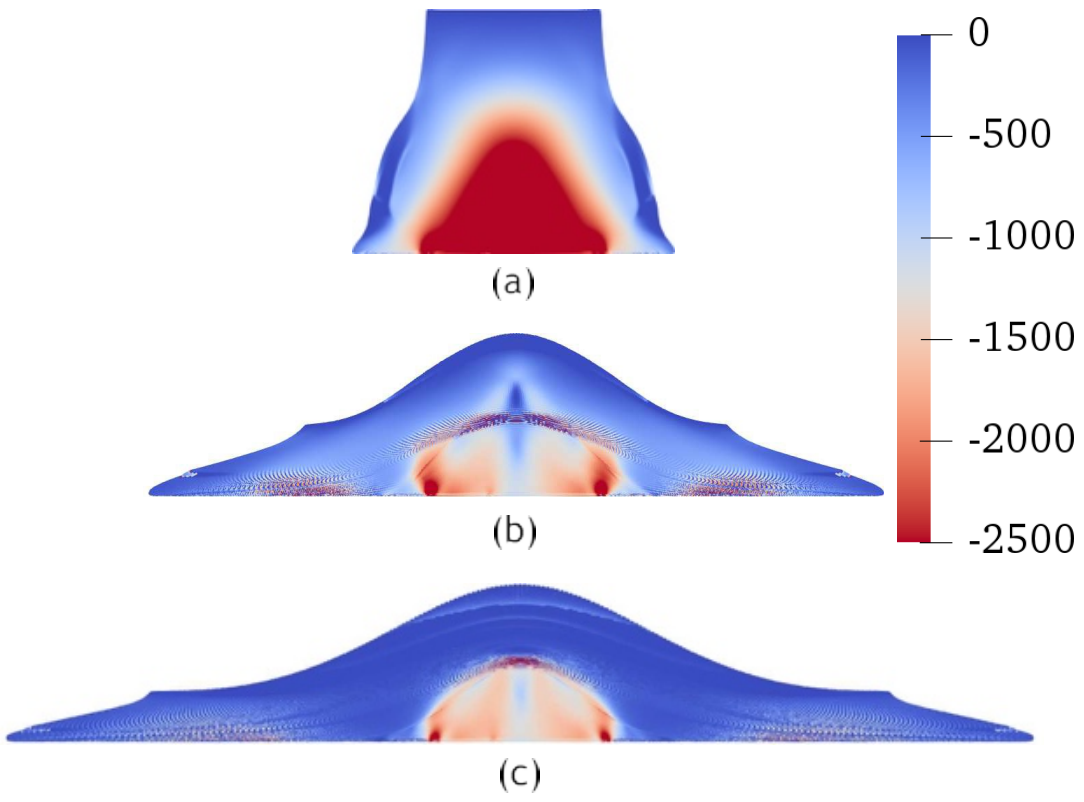

Figure 20: Contours of $\overline{\sigma}_y$ at (a) $t$ = 0.1 s, (b) $t$ = 0.3 s, (c) $t$ = 0.5 s superimposed on the deformed configuration for $a_3 = 0.5$

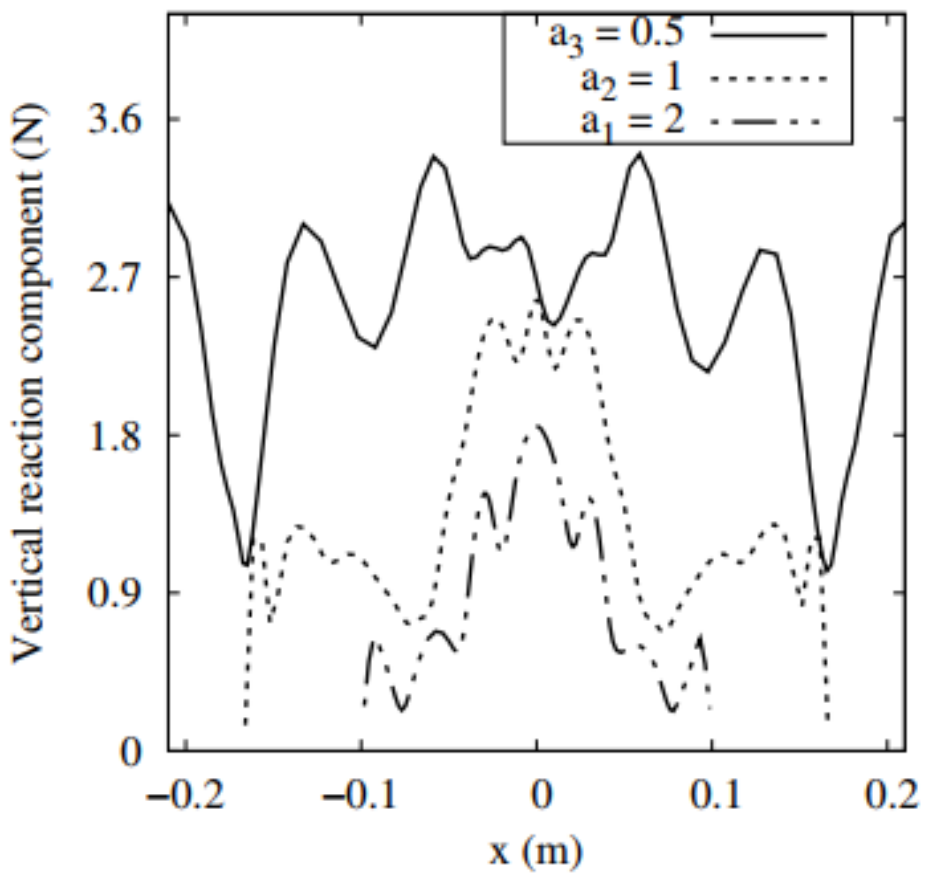

Figure 21: Variation of the magnitude of the vertical reaction force along the base of the column for 3 aspect ratios.

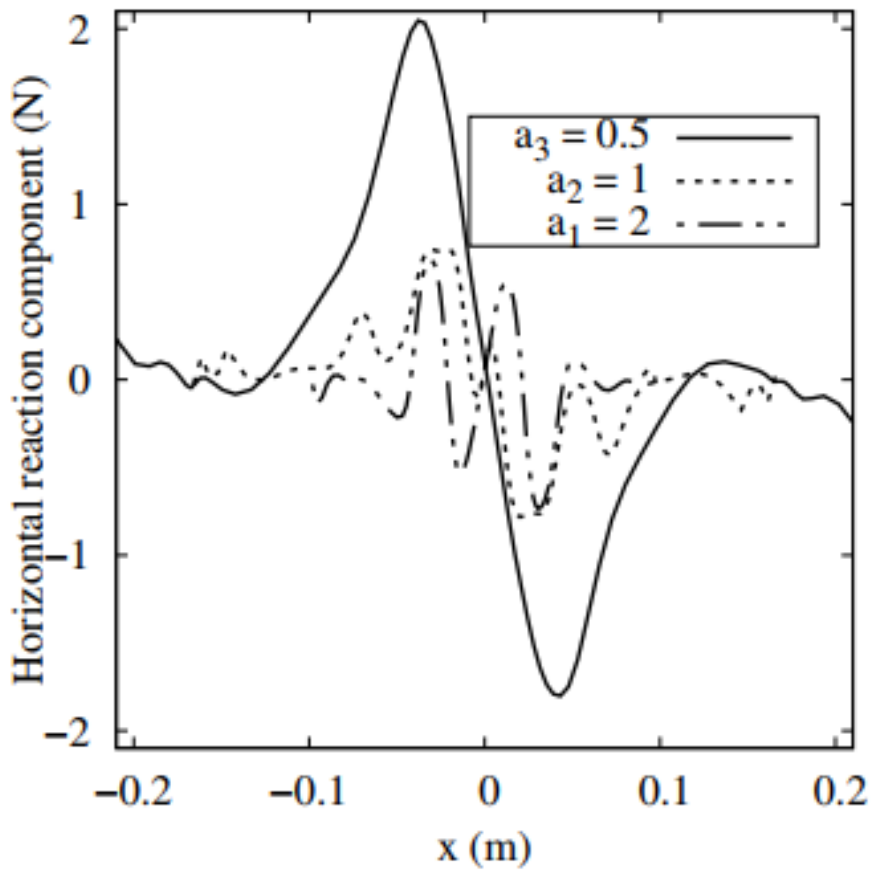

Figure 22: Variation of the magnitude of the horizontal reaction force along the base of the column for 3 aspect ratios.

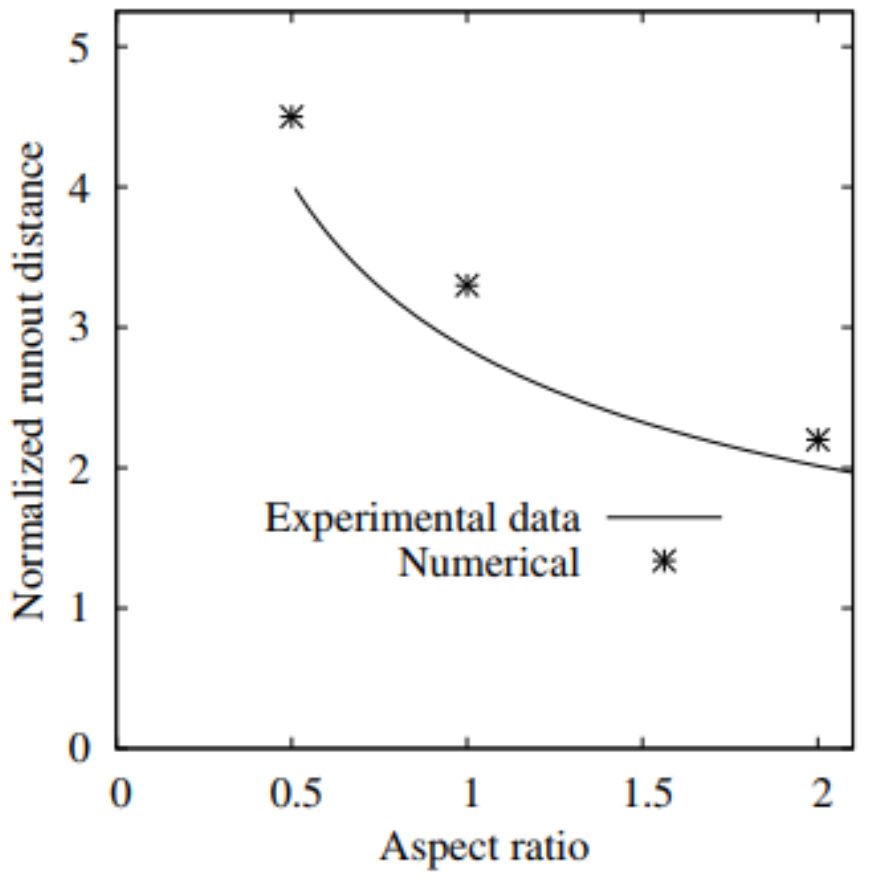

Figure 23: Comparison of the normalized run-out distance with experimental data [37]for different aspect ratios.

*4.2.2. Influence of initial matric suction*

The initial matric suction could have a strong impact on the soil collapse process under unsaturated condition. To test this hypothesis, we run simulations of soil column collapse with three different initial suctions, i.e., $s_1$ = 0 kPa, $s_2$= 25 kPa, and $s_3$ = 50 kPa. The corresponding degrees of saturation $S_r$ = 1.0, 0.92 and 0.88 respectively. The simulations are repeated for the above three aspect ratios. All simulations are conducted under drained conditions with a constant matric suction. The degree of saturation $S_r$ varies with the deformation through the porosity $\varphi$ (i.e., equation (15)).

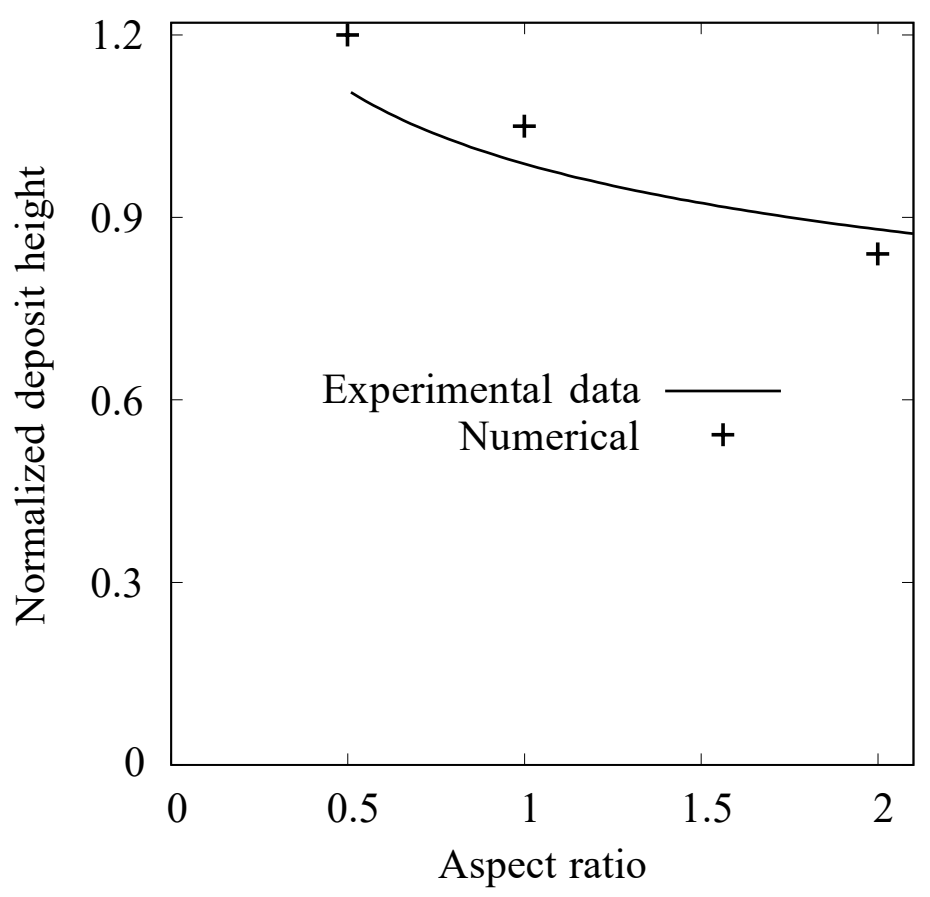

Figure 24: Comparison of the normalized final deposit height with experimental data [37] for different aspect ratios.

The numerical results are presented in Figures 25 - 30. Figures 25, 26, and 27 plot the snapshots of $\varepsilon_s$ in the final deposit configurations from simulations of the soil column with three different initial aspect ratios and matric suctions. The results in Figures 25, 26, and 27 demonstrate that the initial matric suction could significantly affect the final deposit morphology of the soil. The impact of the initial matric suction on the contour of deviatoric strain in the final configuration may depend on the initial aspect ratio of the soil column. For instance, for $a_1$ = 2, as shown in Figure 25 the initial matric suction affects the maximum deviatoric strain in the soil. However, this influence become rather mild when $a_3$ = 0.5 with the same initial column width (i.e., larger initial height of the soil column). Moreover, as shown in Figure 28 the increase of initial matric suction in the soil column reduces the final runout distance for the same aspect ratio. Under the same initial matric suction, the specimen with the larger aspect ratio generates the smaller final runout distance. The final deposit height is generally larger for the specimen with a larger initial matric suction under the same aspect ratio. These observations can be explained by the fact that increasing matric suction generally increases the cohesion of soils (e.g., [1, 55]).

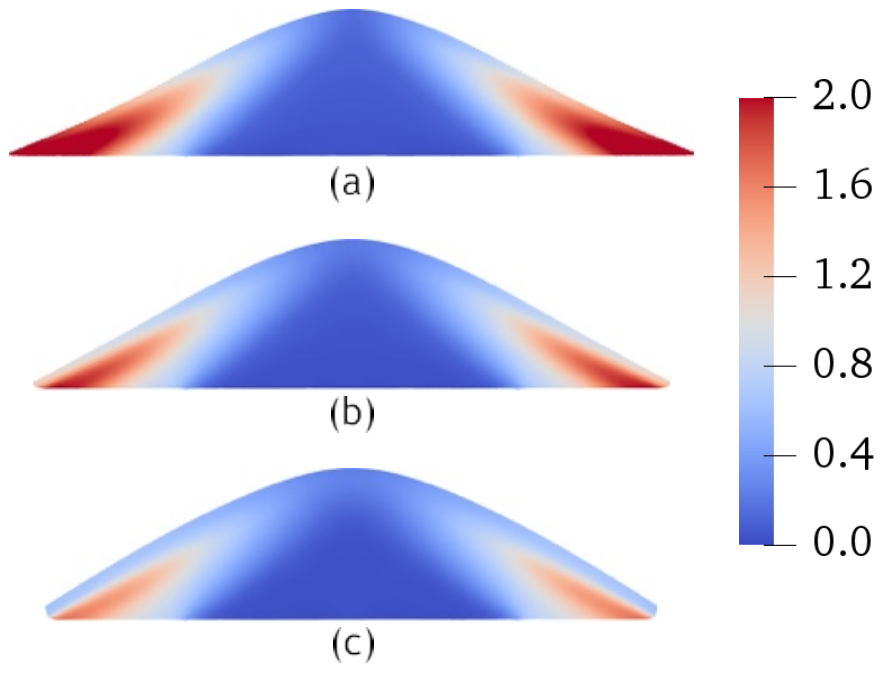

Figure 25: Contours of $\varepsilon_s$ on the final deposit configuration from simulations with (a) $s_1$ = 0 kPa, (b) $s_2$ = 25 kPa, and (c) $s_3$ = 50 kPa for $a_1$ = 2.

To show the sensitivity of shear strains to initial matric suctions, Figure 30 plots $\varepsilon_s$ over time at the selected material points (see Figure 14) in the specimens with different aspect ratios. The results in Figure 30 demonstrate that $\varepsilon_s$ is smaller in the specimen with smaller initial matric suction at the same time for $a_1 = 2$ or $a_2 = 1$. For the specimen with $a_3 = 0.5$ the simulations has the similar shear strains in the early stage of the collapse. However, the equivalent shear strain at the same location becomes larger at the later stage of collapse for the larger initial matric suction. It may be concluded that the impact of initial matric suction on collapse could also depend on aspect ratios.

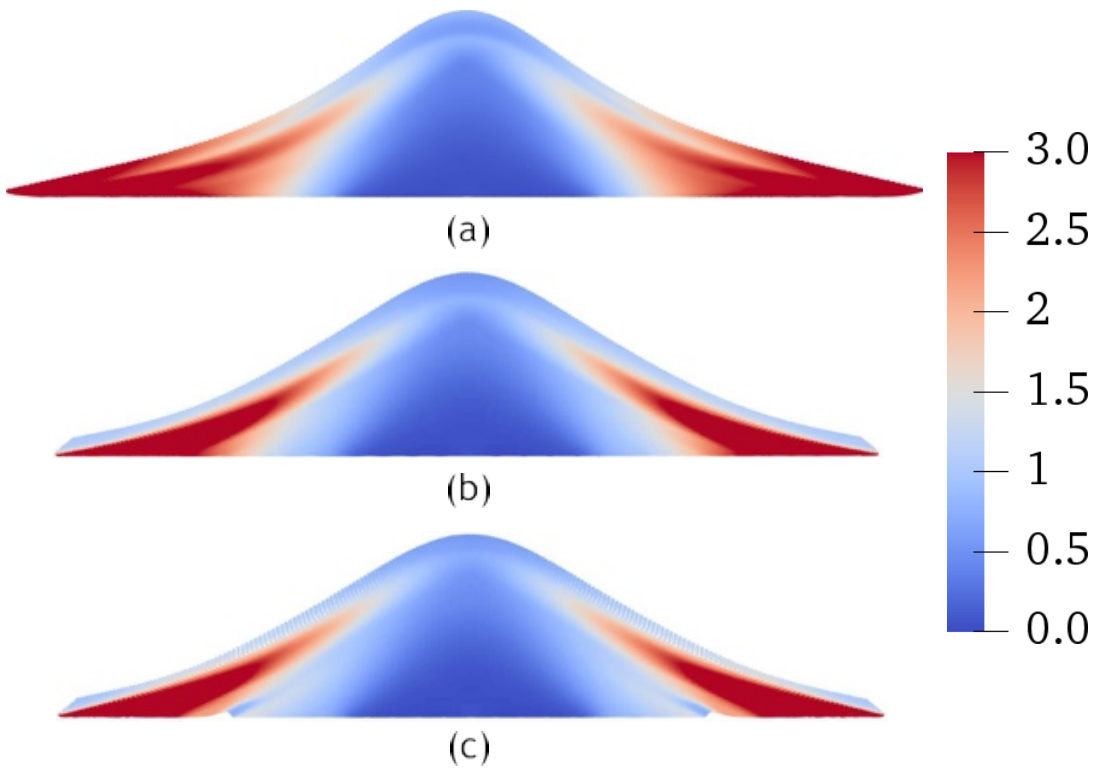

Figure 26: Contours of $\varepsilon_s$ on the final deposit configuration from simulations with (a) $s_1$ = 0 kPa, (b) $s_2$ = 25 kPa, and (c) $s_3$ = 50 kPa for $a_2 = 1$.

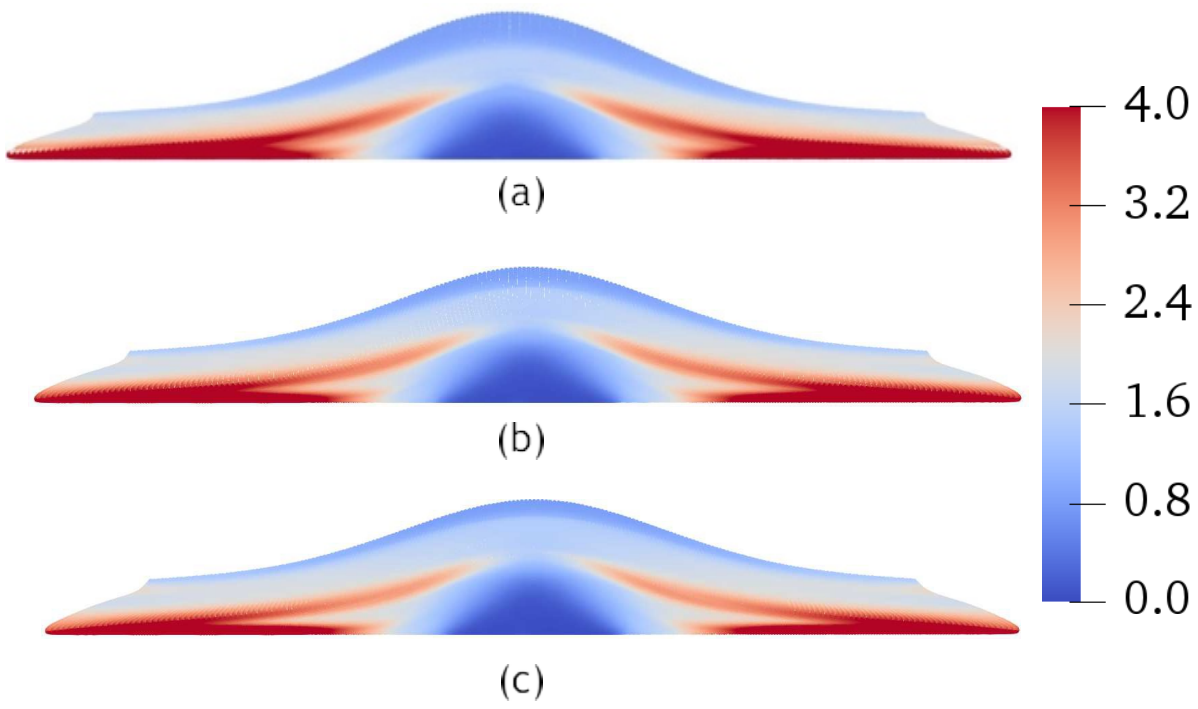

Figure 27: Contours of $\varepsilon_s$ on the final deposit configuration from simulations with (a) $s_1$ = 0 kPa, (b) $s_2$ = 25 kPa, and (c) $s_3$ = 50 kPa for $a_2 = 0.5$.

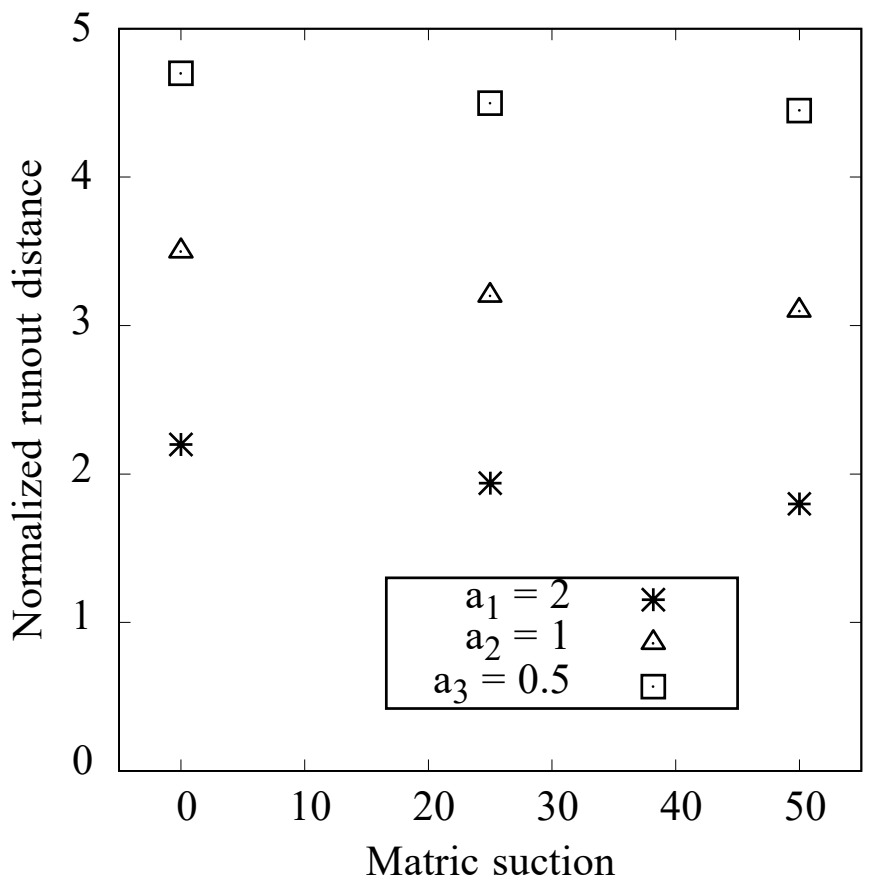

Figure 28: Normalized final runout distance for simulations with different initial matric suctions.

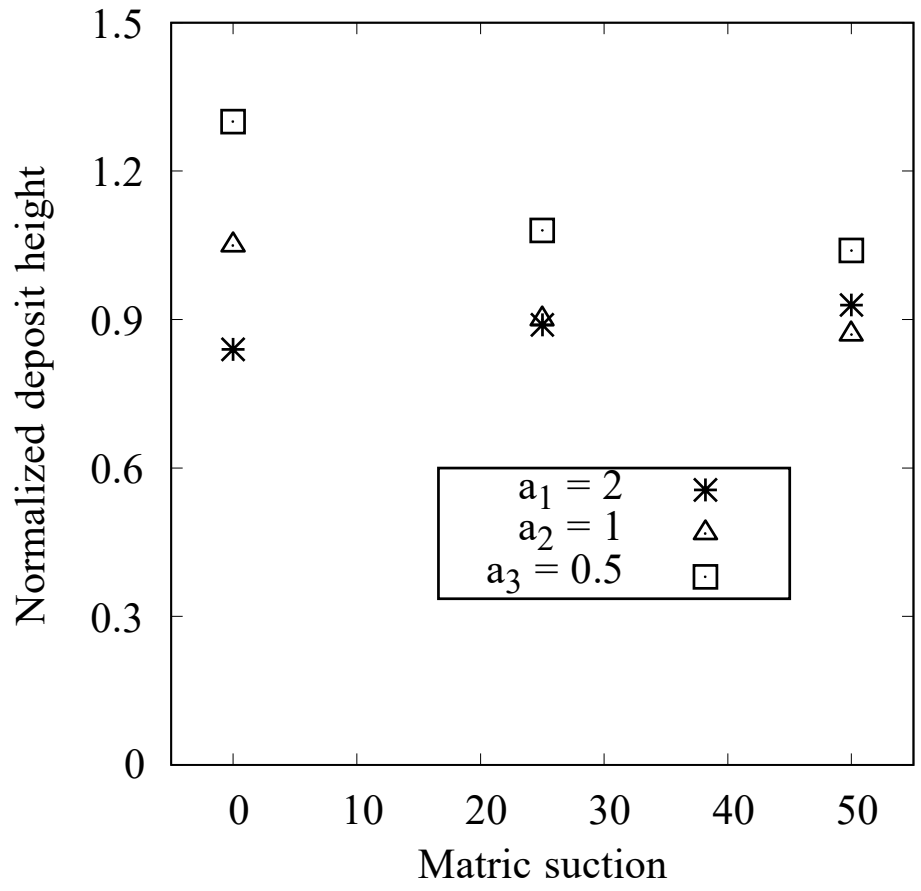

Figure 29: Normalized final deposit height for simulations with different initial matric suctions.

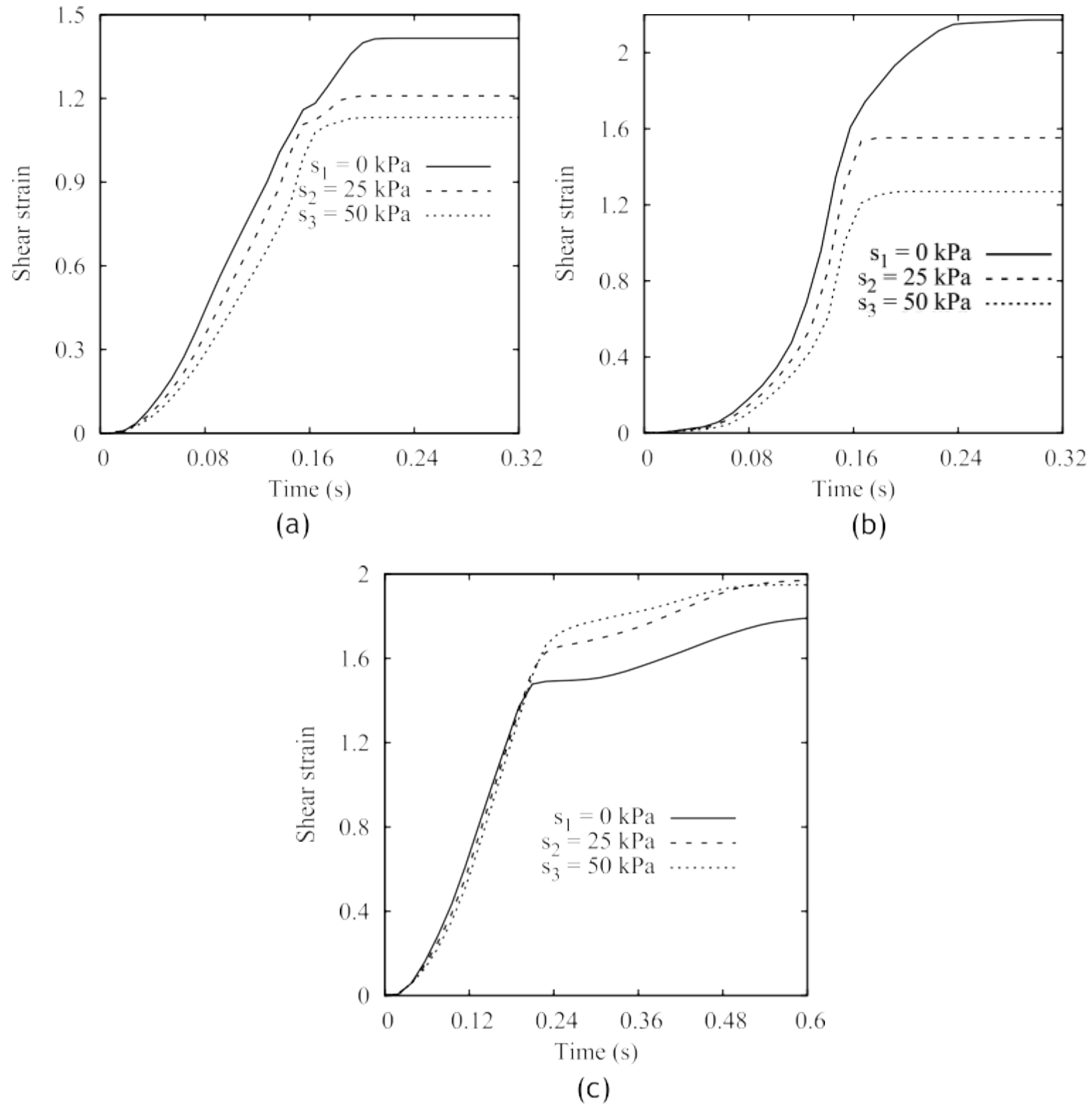

Figure 30: Comparison of $\varepsilon_s$ at the selected locations from simulations using different initial matric suctions with (a) $a_1 = 2$, (b) $a_2 = 1$ and (c) $a_3 = 0.5$.

### *4.2.3. Influence of substrate roughness*

We investigate the influence of substrate roughness on the soil column collapse by running simulations with different friction coefficients of the substrate, i.e., $\mu_{f1} = 0$, $\mu_{f2} = 0.25$, and $\mu_{f3} = 0.5$. The simulations are repeated for the specimens with the above three initial aspect ratios. Figures 31, 32 and 33 present the contours of $\varepsilon_d$ in the final deposit configuration for the simulations with different initial aspect ratios. The results show that the roughness of the substrate could strongly influence the final deposit morphology. This effect increases with smaller aspect ratios. For $a_1 = 2$, the final deposit morphology is in a triangular shape when the substrate is perfectly smooth. Increasing the surface roughness of the substrate causes the final deposit morphology to transition to a more parabolic shape. For $a_2 = 1$ as shown in Figure 32 for a smooth substrate the final deposit morphology has concave slopes. Increasing the substrate roughness changes the final morphology to a more convex shape. For the case of $a_3 = 0.5$ increasing the substrate friction reduces the runout distance of the bottom layer along the substrate while the upper layer of soil has a larger runout distance as $\mu_f$ increases. Figures 34 and 35 present the normalized final runout distance and the deposit height. The results show that the increasing substrate friction could decrease the final runout distance for the specimens with all aspect ratios. For the specimens with larger aspect ratios the deposit height generally increases with larger substrate frictions. However, for the specimen with $a_1 = 2$, the substrate friction has mild influence on the final deposit height as shown in Figure 35.

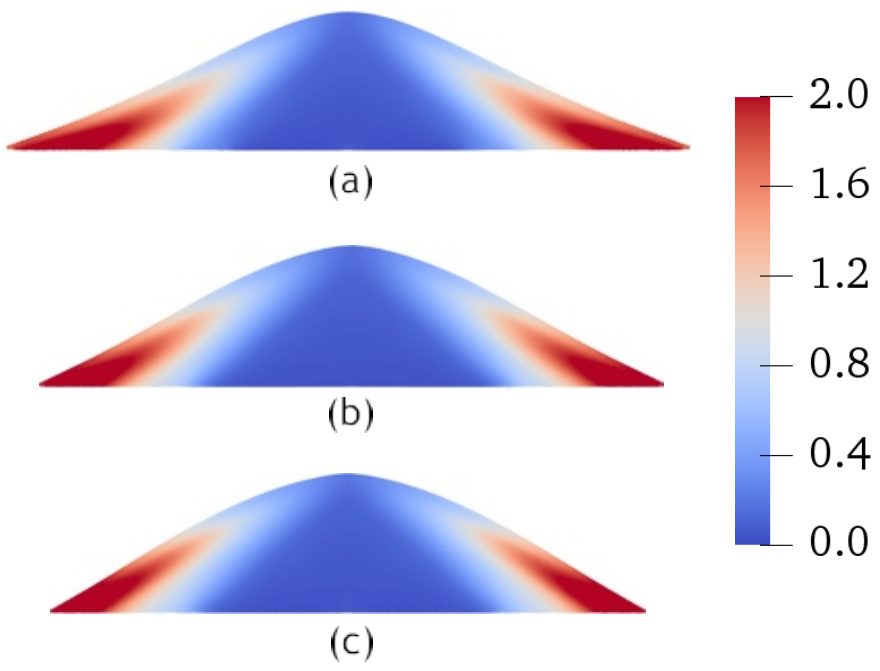

Figure 31: Contours of $\varepsilon_s$ from the simulations using (a) $\mu_{f1} = 0$, (b) $\mu_{f2} = 0.25$, and (c) $\mu_{f3} = 0.5$ for $a_1 = 2$.

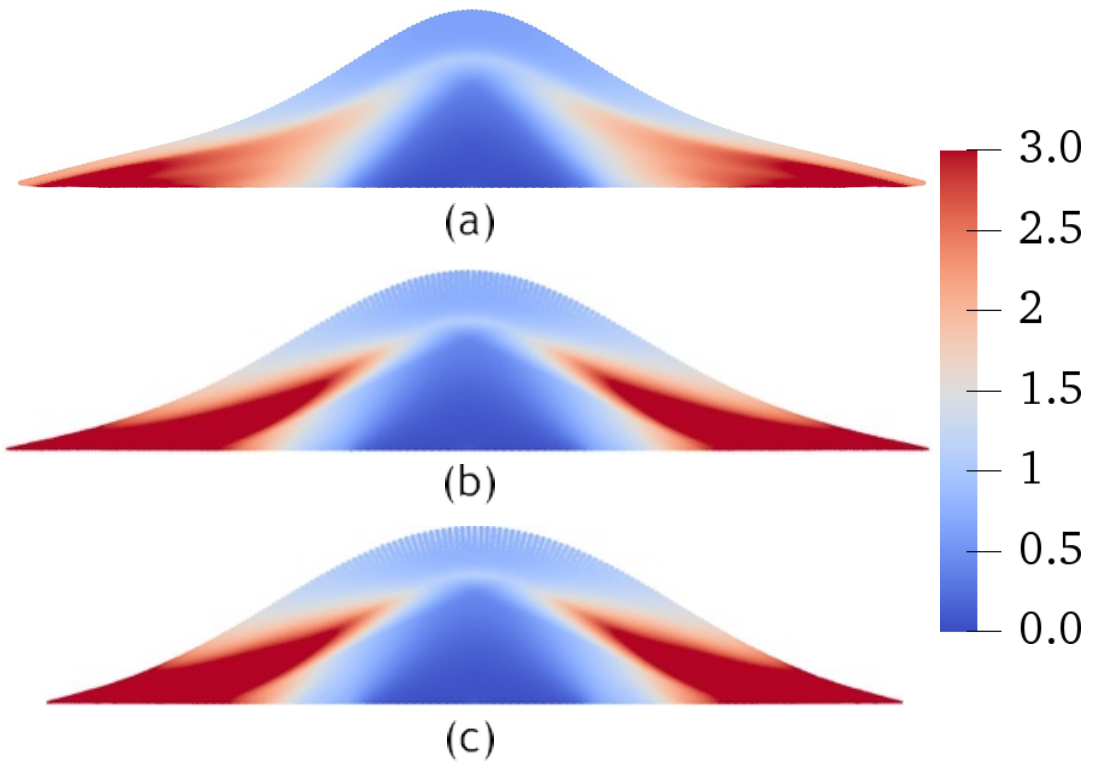

Figure 32: Contours of $\varepsilon_s$ from the simulations using (a) $\mu_{f1} = 0$, (b) $\mu_{f2} = 0.25$, and (c) $\mu_{f3} = 0.5$ for $a_2 = 1$.

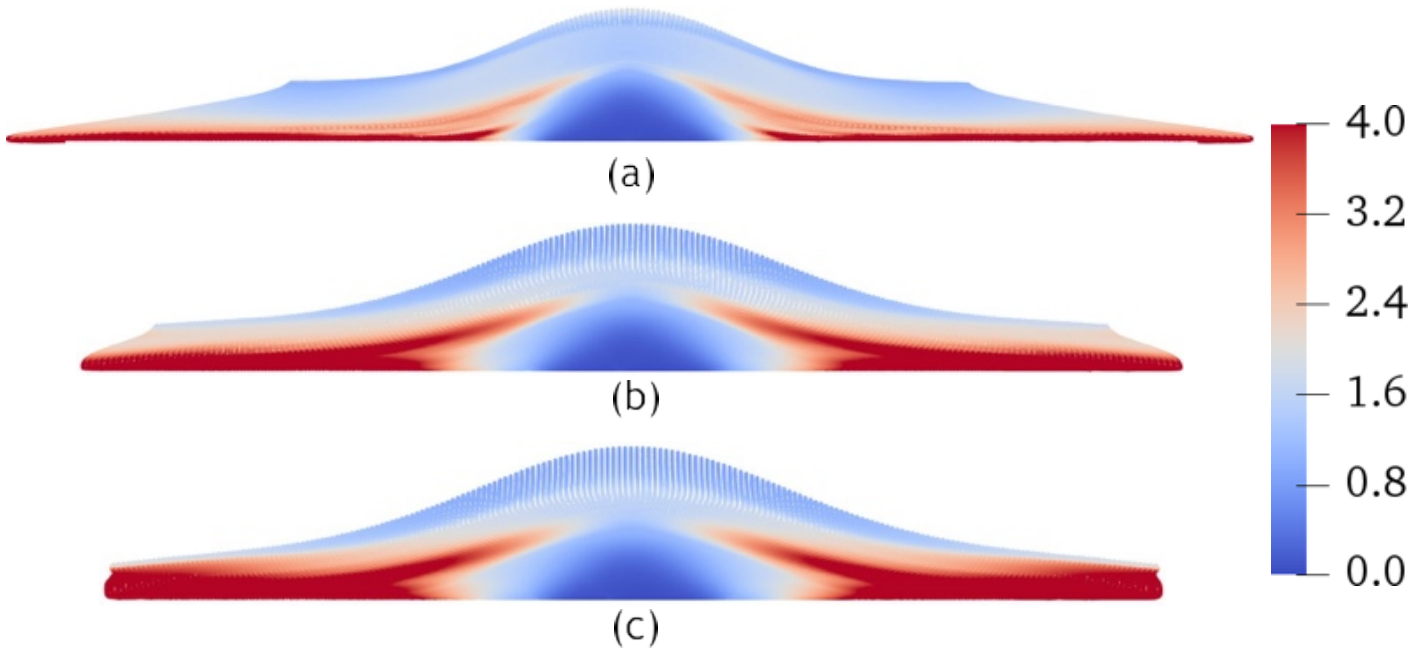

Figure 33: Contours of $\varepsilon_s$ from the simulations using (a) $\mu_{f1} = 0$, (b) $\mu_{f2} = 0.25$, and (c) $\mu_{f3} = 0.5$ for $a_3 = 0.5$.

Finally, Figure 36 compare $\varepsilon_s$ at the selected point from the simulations using different substrate frictions $\mu_f$ (see Figure 14). The results in Figure 36 show that $\varepsilon_s$ at the point is independent of the substrate friction in the early stages of the collapse for all aspect ratios while the substrate friction does affect the maximum shear strain. It could be concluded that the substrate friction influences the final run-out distance and flow behavior but not the collapse triggering mechanism.

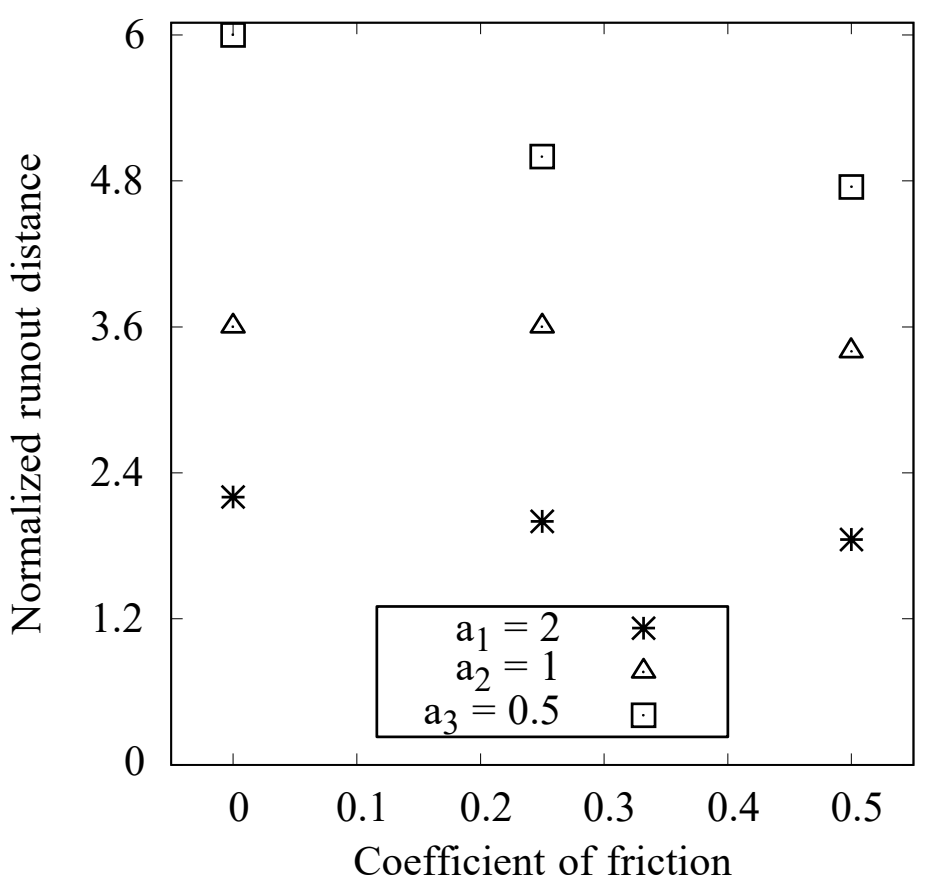

Figure 34: Normalized final runout distance for different substrate frictions.

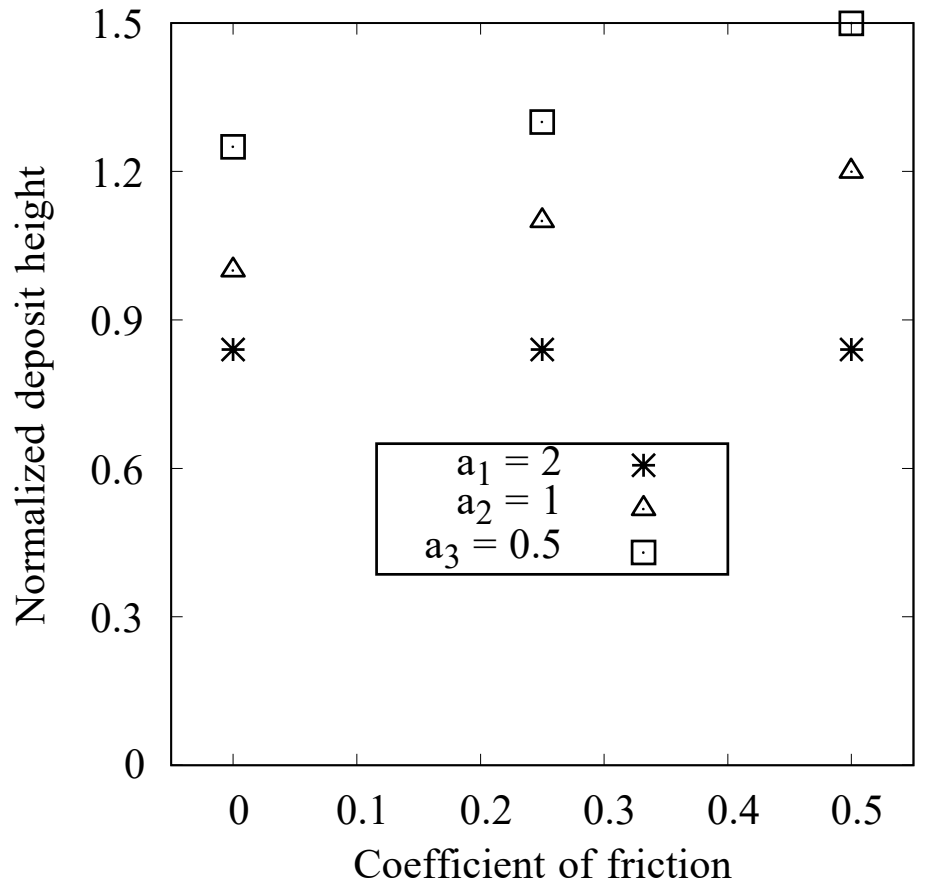

Figure 35: Normalized final deposit height for different substrate frictions.

## 5. Closure

In this article, we have proposed an updated Lagrangian periporomechanics for modeling extreme large deformation in unsaturated porous media under drained conditions (i.e., constant matric suction). In this updated Lagrangian framework it is hypothesized that the family of a material point is a uniform spherical shape (i.e., constant horizon) independent of deformation. In this study the bond-associated sub-horizon concept is utilized to eliminate the zero-energy modes at extreme large deformation of solid skeleton when using the correspondence constitutive models of unsaturated porous media. The stabilized nonlocal velocity gradient in the deformed configuration is used to numerically implement a critical state based visco-plastic model for unsaturated soils. The updated Lagrangian periporomechanics framework is numerically implemented through the explicit Newmark scheme for high-performance computing. The uni-axial compression testing of a rectangular porous material specimen under static and dynamic loads is first presented to demonstrate the stability performance of the updated Lagrangian periporomechanics framework. We then conduct the numerical modeling of soil column collapse to demonstrate the efficacy and robustness of the updated Lagrangian periporomechanics in modeling extreme large deformation in unsaturated porous media under drained conditions. The numerical results have been validated against the experimental data in the literature. We also investigate the influence of initial matric suction and substrate friction on the final deposit morphology. The preliminary numerical results have shown that the impact of initial matric suction on the final run-out distance and deposit height may also depend on the initial aspect ratio.

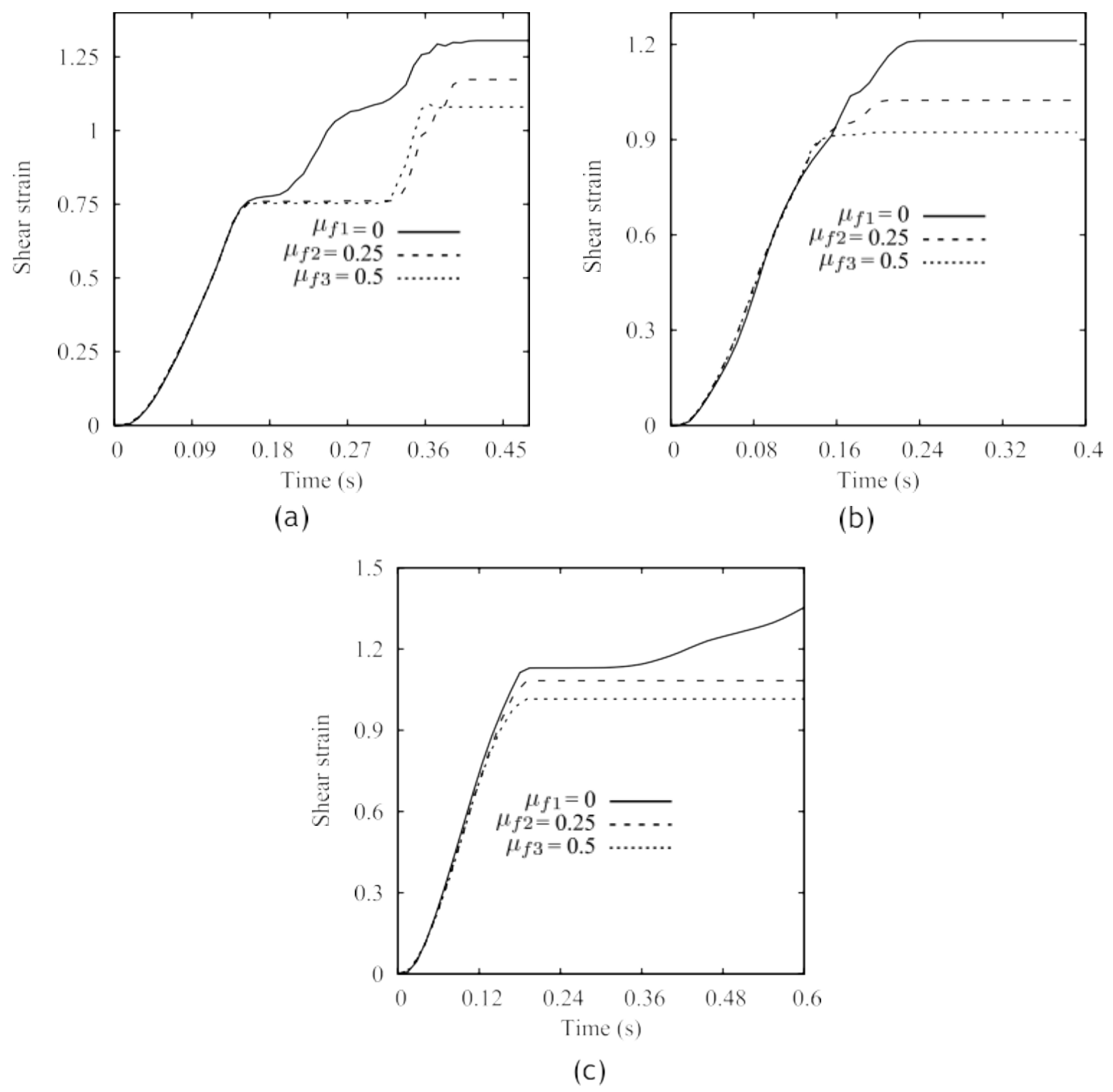

Figure 36: Comparison of $\varepsilon_s$ at the marked point for (a) $a_1 = 2$, (b) $a_2 = 1$ and (c) $a_3 = 0.5$ from simulations using different substrate frictions.

## Acknowledgments

This work has been supported by the US National Science Foundation under contract numbers 1659932 and 1944009. The support is greatly acknowledged.

## Data Availability Statement

The data that support the findings of this study are available from the corresponding author upon reasonable request.

## References

1. Song X, Menon S. Modeling of chemo-hydromechanical behavior of unsaturated porous media: a nonlocal approach based on integral equations. *Acta Geotechnica* 2019;14(3):727–47.
2. Menon S, Song X. A computational periporomechanics model for localized failure in unsaturated porous media. *Computer Methods in Applied Mechanics and Engineering* 2021;384:113932.
3. Song X, Khalili N. A peridynamics model for strain localization analysis of geomaterials. *International Journal for Numerical and Analytical Methods in Geomechanics* 2019;43(1):77–96.
4. Menon S, Song X. Computational multiphase periporomechanics for unguided cracking in unsaturated porous media. *arXiv preprint arXiv:210810433* 2021;.
5. Lewis RW, Lewis RW, Schrefler B. The finite element method in the static and dynamic deformation and consolidation of porous media. John Wiley & Sons; 1998.
6. Zienkiewicz OC, Chan A, Pastor M, Schrefler B, Shiomi T. Computational geomechanics; vol. 613. Citeseer; 1999.
7. Cheng AHD. Poroelasticity; vol. 27. Springer; 2016.

8. Silling S. Reformulation of elasticity theory for discontinuities and long-range forces. *Journal of the Mechanics and Physics of Solids* 2000;48(1):175–209.
9. Silling SA, Epton M, Weckner O, Xu J, Askari E. Peridynamic states and constitutive modeling. *Journal of Elasticity* 2007;88(2):151–84.
10. Song X, Silling SA. On the peridynamic effective force state and multiphase constitutive correspondence principle. *Journal of the Mechanics and Physics of Solids* 2020;145:104161.
11. Menon S, Song X. Shear banding in unsaturated geomaterials through a strong nonlocal hydromechanical model. *European Journal of Environmental and Civil Engineering* 2020;:1–15.
12. Menon S, Song X. A stabilized computational nonlocal poromechanics model for dynamic analysis of saturated porous media. *International Journal for Numerical Methods in Engineering* 2021;.
13. Silling SA, Lehoucq RB. Peridynamic theory of solid mechanics. *Advances in applied mechanics* 2010;44:73–168.
14. Madenci E, Oterkus E. Peridynamic Theory and Its Applications. Springer; 2014.
15. Song X, Borja RI. Mathematical framework for unsaturated flow in the finite deformation range. *International Journal for Numerical Methods in Engineering* 2014;97(9):658–82.
16. Song X, Borja RI. Finite deformation and fluid flow in unsaturated soils with random heterogeneity. *Vadose Zone Journal* 2014;13(5).
17. Bergel GL, Li S. The total and updated lagrangian formulations of state-based peridynamics. *Computational Mechanics* 2016;58(2):351–70.
18. Silling SA, Parks ML, Kamm JR, Weckner O, Rassaian M. Modeling shockwaves and impact phenomena with eulerian peridynamics. *International Journal of Impact Engineering* 2017;107:47–57.
19. Silling SA. Stability of peridynamic correspondence material models and their particle discretizations. *Computer Methods in Applied Mechanics and Engineering* 2017;322:42–57.
20. Silling SA, Askari E. A meshfree method based on the peridynamic model of solid mechanics. *Computers & structures* 2005;83(17-18):1526–35.
21. Breitenfeld M, Geubelle PH, Weckner O, Silling S. Non-ordinary state-based peridynamic analysis of stationary crack problems. *Computer Methods in Applied Mechanics and Engineering* 2014;272:233–50.
22. Tupek MR, Radovitzky R. An extended constitutive correspondence formulation of peridynamics based on nonlinear bond-strain measures. *Journal of the Mechanics and Physics of Solids* 2014;65(1):82–92.
23. Wu CT, Ren B. A stabilized non-ordinary state-based peridynamics for the nonlocal ductile material failure analysis in metal machining process. *Computer Methods in Applied Mechanics and Engineering* 2015;291:197–215.
24. Li P, Hao ZM, Zhen WQ. A stabilized non-ordinary state-based peridynamic model. *Computer Methods in Applied Mechanics and Engineering* 2018;339:262–80.
25. Gu X, Zhang Q, Madenci E, Xia X. Possible causes of numerical oscillations in non-ordinary state-based peridynamics and a bond-associated higher-order stabilized model. *Computer Methods in Applied Mechanics and Engineering* 2019;357:112592.
26. Chen H. Bond-associated deformation gradients for peridynamic correspondence model. *Mechanics Research Communications* 2018;90:34–41.
27. Roy Chowdhury S, Roy P, Roy D, Reddy JN. A modified peridynamics correspondence principle: Removal of zero-energy deformation and other implications. *Computer Methods in Applied Mechanics and Engineering* 2019;346:530–49.
28. Hashim NA, Coombs W, Augarde C, Hattori G. An implicit non-ordinary state-based peridynamics with stabilised correspondence material model for finite deformation analysis. *Computer Methods in Applied Mechanics and Engineering* 2020;371:113304.
29. Chowdhury SR, Roy P, Roy D, Reddy J. A modified peridynamics correspondence principle: Removal of zero-energy deformation and other implications. *Computer Methods in Applied Mechanics and Engineering* 2019;346:530–49.
30. Gabriel E, Fagg GE, Bosilca G, Angskun T, Dongarra JJ, Squyres JM, Sahay V, Kambadur P, Barrett B, Lumsdaine A, et al. Open mpi: Goals, concept, and design of a next generation mpi implementation. In: *European Parallel Virtual Machine/Message Passing Interface Users' Group Meeting*. Springer; 2004:97–104.
31. Utili S, Zhao T, Houlsby G. 3d dem investigation of granular column collapse: evaluation of debris motion and its destructive power. *Engineering geology* 2015;186:3–16.
32. Lajeunesse E, Quantin C, Allemand P, Delacourt C. New insights on the runout of large landslides in the valles-marineris canyons, mars. *Geophysical Research Letters* 2006;33(4).
33. Alonso EE. Triggering and motion of landslides. *Géotechnique* 2021;71(1):3–59.
34. Krantz M, Zhang H, Zhu J. Characterization of powder flow: Static and dynamic testing. *Powder Technology* 2009;194(3):239–45.
35. Trolese M, Cerminara M, Ongaro TE, Giordano G. The footprint of column collapse regimes on pyroclastic flow temperatures and plume heights. *Nature communications* 2019;10(1):1–10.
36. Lajeunesse E, Mangeney-Castelnau A, Vilotte JP. Spreading of a granular mass on a horizontal plane. *Physics of Fluids* 2004;16(7):2371–81.
37. Lube G, Huppert HE, Sparks RSJ, Hallworth MA. Axisymmetric collapses of granular columns. *Journal of Fluid Mechanics* 2004;508:175–99.
38. Crosta G, Imposimato S, Roddeman D. Numerical modeling of 2-d granular step collapse on erodible and nonerodible surface. *Journal of Geophysical Research: Earth Surface* 2009;114(F3).
39. Bui HH, Fukagawa R, Sako K, Ohno S. Lagrangian meshfree particles method (sph) for large deformation and failure flows of geomaterial using elastic–plastic soil constitutive model. *International journal for numerical and analytical methods in geomechanics* 2008;32(12):1537–70.
40. Dunatunga S, Kamrin K. Continuum modelling and simulation of granular flows through their many

phases. *J Fluid Mech* 2015;779:483–513.
41. Mast CM, Arduino P, Mackenzie-Helnwein P, Miller GR. Simulating granular column collapse using the Material Point Method. *Acta Geotechnica* 2015;10(1):101–16.
42. Ni T, Pesavento F, Zaccariotto M, Galvanetto U, Zhu QZ, Schrefler BA. Hybrid fem and peridynamic simulation of hydraulic fracture propagation in saturated porous media. *Computer Methods in Applied Mechanics and Engineering* 2020;366:113101.
43. Ni T, Pesavento F, Zaccariotto M, Galvanetto U, Schrefler BA. Numerical simulation of forerunning fracture in saturated porous solids with hybrid fem/peridynamic model. *Computers and Geotechnics* 2021;133:104024.
44. Menon S, Song X. A computational periporomechanics model for localized failure in unsaturated porous media. *arXiv preprint arXiv:201015793* 2020;.
45. Song X, Wang K, Bate B. A hierarchical thermo-hydro-plastic constitutive model for unsaturated soils and its numerical implementation. *International Journal for Numerical and Analytical Methods in Geomechanics* 2018;42(15):1785–805.
46. Cao J, Jung J, Song X, Bate B. On the soil water characteristic curves of poorly graded granular materials in aqueous polymer solutions. *Acta Geotechnica* 2018;13(1):103–16.
47. Niu WJ, Ye WM, Song X. Unsaturated permeability of gaomiaozi bentonite under partially free-swelling conditions. *Acta Geotechnica* 2020;15(5):1095–124.
48. Song X. Transient bifurcation condition of partially saturated porous media at finite strain. *International Journal for Numerical and Analytical Methods in Geomechanics* 2017;41(1):135–56.
49. Simo JC, Hughes TJ. Computational inelasticity; vol. 7. Springer Science & Business Media; 1998.
50. Borja RI. Cam-clay plasticity. part v: A mathematical framework for three-phase deformation and strain localization analyses of partially saturated porous media. *Computer methods in applied mechanics and engineering* 2004;193(48-51):5301–38.
51. Hughes TJ. The finite element method: linear static and dynamic finite element analysis. Courier Corporation; 2012.
52. Littlewood DJ, Shelton T, Thomas JD. Estimation of the critical time step for peridynamic models. Tech. Rep.; Sandia National Lab.(SNL-NM), Albuquerque, NM (United States); 2013.
53. Belytschko T, Liu WK, Moran B, Elkhodary K. Nonlinear finite elements for continua and structures. John wiley & sons; 2014.
54. Silling SA. Meshfree peridynamics for soft materials. Tech. Rep.; Sandia National Lab.(SNL-NM), Albuquerque, NM (United States); 2016.
55. Wang K, Song X. Strain localization in non-isothermal unsaturated porous media considering material heterogeneity with stabilized mixed finite elements. *Computer Methods in Applied Mechanics and Engineering* 2020;359:112770.